\documentclass[12pt, draftclsnofoot, onecolumn]{IEEEtran}
\usepackage{amsmath,amssymb,dsfont,stfloats,color,url,bbm}
\usepackage[pdftex]{graphicx}
\usepackage{subfigure}
\usepackage{cite}
\usepackage{xcolor}
\allowdisplaybreaks

\DeclareGraphicsExtensions{.eps,.pdf,.png,.jpg,.gif,.jpeg,.pstex}

\newtheorem{remark}{Remark}
\newtheorem{example}{Example}

\newfont{\bbb}{msbm10 scaled 700}

\newfont{\bb}{msbm10 scaled 1100}
\newcommand{\CC}{\mbox{\bb C}}
\newcommand{\PP}{\mbox{\bb P}}

\newcommand{\EE}{\mbox{\bb E}}

\newcommand{\HH}{\mbox{\bb H}}
\newcommand{\SSS}{\mbox{\bb S}}
\newcommand{\UU}{\mbox{\bb U}}

\newcommand{\BB}{\mbox{\bb B}}
\renewcommand{\AA}{\mbox{\bb A}}

\newcommand{\yy}{\mathbbm{y}}
\newcommand{\xx}{\mathbbm{x}}
\newcommand{\zz}{\mathbbm{z}}
\newcommand{\sss}{\mathbbm{s}}

\newcommand{\hh}{\mathbbm{h}}
\newcommand{\uu}{\mathbbm{u}}
\newcommand{\vvv}{\mathbbm{v}}

\newcommand{\av}{{\bf a}}

\newcommand{\ev}{{\bf e}}

\newcommand{\hv}{{\bf h}}

\newcommand{\nv}{{\bf n}}

\newcommand{\qv}{{\bf q}}
\newcommand{\rv}{{\bf r}}
\newcommand{\sv}{{\bf s}}

\newcommand{\uv}{{\bf u}}
\newcommand{\wv}{{\bf w}}
\newcommand{\vv}{{\bf v}}

\newcommand{\yv}{{\bf y}}
\newcommand{\zv}{{\bf z}}
\newcommand{\zerov}{{\bf 0}}

\newcommand{\Am}{{\bf A}}
\newcommand{\Bm}{{\bf B}}

\newcommand{\Dm}{{\bf D}}
\newcommand{\Em}{{\bf E}}
\newcommand{\Fm}{{\bf F}}
\newcommand{\Gm}{{\bf G}}
\newcommand{\Hm}{{\bf H}}
\newcommand{\Id}{{\bf I}}

\newcommand{\Nm}{{\bf N}}

\newcommand{\Sm}{{\bf S}}

\newcommand{\Um}{{\bf U}}

\newcommand{\Vm}{{\bf V}}

\newcommand{\Ym}{{\bf Y}}
\newcommand{\Zm}{{\bf Z}}

\newcommand{\Cc}{{\cal C}}

\newcommand{\Ec}{{\cal E}}

\newcommand{\Gc}{{\cal G}}

\newcommand{\Ic}{{\cal I}}

\newcommand{\Lc}{{\cal L}}

\newcommand{\Nc}{{\cal N}}

\newcommand{\Sc}{{\cal S}}

\newcommand{\Uc}{{\cal U}}

\newcommand{\nuv}{\hbox{\boldmath$\nu$}}
\newcommand{\muv}{\hbox{\boldmath$\mu$}}
\newcommand{\zetav}{\hbox{\boldmath$\zeta$}}
\newcommand{\phiv}{\hbox{\boldmath$\phi$}}

\newcommand{\xiv}{\hbox{\boldmath$\xi$}}

\newcommand{\Gammam}{\hbox{\boldmath$\Gamma$}}

\newcommand{\Sigmam}{\hbox{\boldmath$\Sigma$}}

\newcommand{\Thetam}{\hbox{\boldmath$\Theta$}}

\newcommand{\diag}{{\hbox{diag}}}

\newcommand{\trace}{{\hbox{tr}}}

\renewcommand{\arg}{{\hbox{arg}}}

\newcommand{\eqdef}{\stackrel{\Delta}{=}}

\newcommand{\herm}{{\sf H}}

\newcommand{\SINR}{{\sf SINR}}
\newcommand{\SNR}{{\sf SNR}}
\newcommand{\MMSE}{{\sf MMSE}}

\newcommand{\taudmrs}{\tau_p}

\usepackage{stmaryrd} 

\newcommand{\BLUE}{\color[rgb]{0,0,0.90}}

\usepackage{hyperref}
\hypersetup{
    bookmarks=true,         
    unicode=false,          
    pdftoolbar=true,        
    pdfmenubar=true,        
    pdffitwindow=false,     
    pdfstartview={FitH},    
    pdfnewwindow=true,      
    colorlinks=true,       
    linkcolor=red,          
    citecolor=green,        
    filecolor=blue,      
    urlcolor=blue           
}

\usepackage{siunitx}

\renewcommand{\arg}{{\rm arg}}
\newcommand*{\Resize}[2]{\resizebox{#1}{!}{$#2$}}%

\title{Subspace-Based Pilot
	Decontamination in User-Centric Scalable Cell-Free
	Wireless Networks}

\begin{document}
	\bstctlcite{BSTcontrol}

	\author{\IEEEauthorblockN{F. G\"ottsch\IEEEauthorrefmark{1},
			N. Osawa\IEEEauthorrefmark{2}, T. Ohseki\IEEEauthorrefmark{2}, K. Yamazaki\IEEEauthorrefmark{2}, G. Caire\IEEEauthorrefmark{1}}
		\thanks{\IEEEauthorrefmark{1}Technical University of Berlin, Germany. 
			\IEEEauthorrefmark{2}KDDI Research Inc., Japan.
	}}
	
	\maketitle

	
	
	
	
	
	
	
	
	
	\begin{abstract}
		We consider a cell-free wireless system operated in Time Division Duplex (TDD) mode with user-centric clusters of remote radio units (RUs). 
		Since the uplink pilot dimensions per channel coherence slot is limited, co-pilot users might incur mutual pilot contamination.
		In the current literature, it is assumed that the long-term statistical knowledge of all user channels is available. This enables Minimum Mean-Square Error channel estimation or simplified
		dominant subspace projection, which achieves significant pilot decontamination  under certain assumptions on the channel covariance matrices. 
		However, estimating the channel covariance matrix or even just its dominant subspace at all RUs forming a user cluster is not 
		an easy task. In fact, if not properly designed, a piloting scheme for such long-term statistics estimation will also be subject to the contamination problem. 
		In this paper, we propose a new channel subspace estimation scheme explicitly designed for cell-free wireless networks. Our scheme is based on 1)  
		a sounding reference signal (SRS)  using  latin squares wideband frequency hopping, and 2)  a subspace estimation method based on 
		robust Principal Component Analysis (R-PCA).  The SRS hopping scheme ensures that for any user and any RU participating in its cluster, only a few pilot measurements will contain strong co-pilot interference. These few heavily contaminated measurements are (implicitly) eliminated by R-PCA, which is designed to regularize the estimation and discount the ``outlier'' measurements. 
		Our simulation results show that the proposed scheme achieves almost perfect subspace knowledge, which in turns  yields system performance
		very close to that with ideal channel state information, thus essentially solving the problem of pilot contamination in cell-free user-centric TDD wireless networks. 
	\end{abstract} 
	
	\begin{IEEEkeywords}
		User-centric, cell-free wireless networks, pilot decontamination, principal component analysis
	\end{IEEEkeywords}
	
	\section{Introduction}  \label{sec:intro}
	
	Multiuser Multiple-Input Multiple-Output (MIMO) has been widely investigated from a theoretical viewpoint \cite{Caire-Shamai-TIT03,Viswanath-Tse-TIT03,Weingarten-Steinberg-Shamai-TIT06,Caire-Jindal-Kobayashi-Ravindran-TIT10} and has become
	a very important component of highly spectrally efficient physical layer (PHY) design in cellular \cite{3gpp38211,Larsson-book,9336188} and local area wireless networks (see  \cite{khorov2018tutorial,qu2019survey} and references therein). 
	A successful related concept is massive MIMO \cite{marzetta2010noncooperative}, where the number of base station (BS) antennas $M$ 
	is much larger than the number $K$ of single antenna user equipments (UEs) served simultaneously (i.e., in the same time-frequency slot) 
	by spatial multiplexing. 
	A key idea in massive MIMO \cite{marzetta2010noncooperative,Larsson-book} is that, 
	thanks to the reciprocity of the uplink (UL) and downlink (DL) channels achieved by Time Division Duplex (TDD) operations,\footnote{This reciprocity condition
		is verified if the UL and DL slots occur within an interval significantly shorter than a channel coherence time, and if the Tx/Rx hardware of the BS radio are
		calibrated. Hardware calibration has been widely demonstrated in practical testbeds, and can be achieved either using particular RF design solutions 
		(e.g., see \cite{benzin2017internal}) or using over-the-air calibration (e.g., see \cite{rogalin2014scalable}). 
		As far as the channel coherence time is concerned, a typical mobile user 
		at carrier frequency between 2.0 and 3.7 GHz incurs Doppler typical spreads of $\sim 100$ Hz corresponding to channel coherence times of $\sim 10$ ms.  
		For example, with TDD slots of 1 ms (corresponding to a subframe of a 10ms frame of 5G new radio (5GNR)), 
		we are well in the range for which the channel can be considered time-invariant 
		over a UL/DL cycle.
		For faster moving users or higher carrier frequencies, the phenomenon of ``channel aging'' \cite{truong2013effects} between the UL and the DL slot 
		cannot be neglected any longer. This however goes beyond the scope of the present paper.}
	an arbitrarily large number $M$ of BS antennas can be trained by a finite number of $K$ UEs
	using only a  finite-dimensional UL pilot signal with dimension $\taudmrs \geq K$. 
	
	As a further extension of the concept of Multiuser MIMO, joint processing of spatially distributed remote radio units (RUs), which can be traced back to \cite{wyner1994shannon}, was investigated in different contexts and under different names such as {\em coordinate multipoint}, {\em cloud radio access network}, or {\em cell-free massive MIMO} \cite{7827017, 7917284, 8845768}.  In this paper we focus on {\em scalable user-centric architectures} as defined in \cite{9336188,9064545}, where each UE is served by a finite size cluster 
	of RUs and, in turn, each RU serves a finite number of UEs. Here, scalability is defined as the property that 
	the data rate at any network node remains finite as the network area $A$ grows to infinity with a given constant density of UEs and RUs. In particular, we can assume that cluster processing is performed at given decentralized units (DUs) whose density is also constant, such that 
	each DU processes only a finite number of clusters.  Non-cooperative traditional cellular networks \cite{marzetta2010noncooperative} are of 
	course a special case of such class of scalable systems, where clusters have size 1 (cells) and are (logically) co-located with DUs. 
	User-centric cell-free massive MIMO is seen as one of the promising approaches to serve simultaneously a large number of UEs with uniformly high service quality in dense 5G wireless networks, as the effects of pathloss and blocking are reduced and macro-diversity is facilitated due to the more uniform 
	spatial distribution of the RU antennas with respect to a conventional cellular massive MIMO system. 
	
	As the network size becomes large, the total number of UEs in the system will eventually become (much) larger than the dimension $\taudmrs$
	of the UL demodulation reference signal (DMRS) for channel estimation. In this case, the phenomenon known as 
	{\em pilot contamination} will appear. In short, when one RU receives more than one user sending the same UL DMRS pilot, it will estimate a linear combination of their channels. Using this estimate to calculate the UL linear detector and the DL linear precoder leads to a coherent interference term that does not disappear
	even in the limit of large number of RU antennas per user \cite{marzetta2010noncooperative}. While such effect can be mitigated in a conventional cellular system
	by allocating orthogonal pilots inside each cell, the problem is further exacerbated in cell-free user-centric networks where the concept of ``cell'' does not
	strictly exist, and the user-centric clusters are generally intertwined.

	\subsection{Related works}
	
	In our previous work \cite{goettsch2021impact}, in order to reflect the localized and distributed nature of the user-centric cluster processing in a simple way, 
	we have considered the case where each RU can obtain channel state information (CSI) only from the UL pilots of its  
	associated UEs.  The case where this information is perfect is referred to as {\em ideal partial CSI}. In practice,  the CSI must be extracted from the (noisy and contaminated) UL pilot signals. Hence, we proposed a pilot decontamination/estimation scheme
	based on projecting the UL pilot observation onto the dominant subspace of the channel covariance matrix, where such information, pertaining to the channel 
	long-term statistics \cite{adhikary2013joint,haghighatshoar2017massive,9336188}, was assumed to be perfectly known. 
	The proposed subspace projection method in \cite{goettsch2021impact} was shown to essentially recover the performance of ideal partial CSI both for the UL and for 
	the DL (see also  \cite{kddi_uldl_precoding}).

		It should be noticed here that a large number of works on cell-free user-centric wireless networks
		makes (implicitly or explicitly) the assumption that the channel long-term statistics are known a priori, with the justification that, 
		since such statistics change slowly in time and are constant with frequency
		over the signal bandwidth,\footnote{The invariance of the channel spatial correlation over the signal bandwidth 
		is well-known and holds true under the so-called ``narrowband'' assumption, i.e., when the signal bandwidth is
		significantly (less than 10\%) of the carrier frequency. This is indeed the case in most wireless communication systems. For example, 5G systems in the sub-6GHz frequency tier have a signal bandwidth generally $\leq 100$MHz and 
		carrier frequency $\geq 2$ GHz.} 
		they can be ``easily'' estimated (e.g., see \cite{9064545,9336188,8845768} and references therein). In these works, 
		Minimum Mean-Square Error (MMSE) channel estimation is applied to the contaminated UL pilot received signal under the assumption of 
		known channel covariance matrices. Due to the fact that the channel subspace of the desired user and that spanned by the co-pilot interference are typically not identical, the MMSE linear projection yields a ``decontamination'' effect
		very similar to the subspace projection method proposed in 
		\cite{goettsch2021impact}.\footnote{In fact, the subspace projection method can be seen as a low-complexity approximation of the linear MMSE estimator in the case of high SNR.}
		Therefore, all these previous works are characterized by the same problem, namely,  how to obtain ``clean'' 
		UL channel measurements from all UEs associated to a given RU and in number 
		sufficient to have a good estimate of the desired user channel covariance matrix, in the time span of a few seconds during which the propagation geometry can be considered stationary (notice that on longer time scales the channel covariance matrix would also be time-variant). 
	
	\subsection{Contributions}
	
	 In this paper, we address the above problem by proposing a new explicit scheme to efficiently obtain 
		the channel dominant subspace for each UE and all RUs forming its user-centric cluster.
	We note that the 5GNR standard specifies two types of UL pilots, namely, 
	the DMRS and the sounding reference signal (SRS) \cite{3gpp38211}. 
	We propose specific DMRS and SRS pilot assignment schemes for the instantaneous channel and the (long-term) subspace estimation, respectively. 
	While the usage of DMRS pilots is standard in the literature, the introduction of SRS pilots for subspace estimation in cell-free wireless systems is novel.  The 
		proposed SRS pilot scheme consists of assigning 
		a time-frequency hopping sequence to each UE based on a set of 
		orthogonal  latin squares \cite{467960} in order to randomize the  pilot contamination in the received SRS sequence. 
	Since the channel covariance matrix is invariant with respect to the subcarrier index over the whole signal bandwidth, the SRS hopping scheme 
	is ``wideband'', i.e., it is not limited to an individual slot (or resource block) formed by a small set of adjacent subcarriers. In contrast, 
	each sequence hops over all the subcarriers spanning the whole channel bandwidth. 
	The orthogonal latin square hopping assignment guarantees that, with high probability, only a small number of highly contaminated SRS measurements are collected at each RU for its associated users.  
	 It is important to notice here that simple averaging over several time slots is not sufficient to eliminate such contamination. For example, suppose that a single SRS pilot sample is affected by strong co-pilot interference 
		20 dB above the signal of the desired user, due to the near-far effect. Then, 
		more than 100 samples would be necessary to ``average out'' such strong contaminated sample.
	This means that the channel subspace obtained from standard Principal Component Analysis (PCA) is generally very significantly degraded by the presence of a few highly contaminated ``outliers'' in the SRS pilots received samples. 
	In order to cope with these strong but sparse {\em measurement outliers}, we propose the use of a robust PCA (R-PCA) algorithm to extract the subspace information from the SRS pilots. 
	Several R-PCA algorithms are well-known in the signal processing literature. However, the proposal to use such schemes in the context of channel subspace estimation for cell-free user-centric networks is novel. 
	Furthermore, owing to the fact that for uniform linear arrays (ULAs) and uniform planar arrays (UPAs), as widely used in today's massive MIMO implementations, 
	the channel covariance matrix is Toeplitz (for ULA) or Block-Toeplitz (for UPA), and that large Toeplitz and block-Toeplitz matrices are approximately diagonalized by
	discrete Fourier transforms (DFT) on the columns and on the rows
	(see  \cite{adhikary2013joint} for a precise statement based on Szeg\"o's theorem), we further {\rm rectify} the R-PCA by projecting the estimated subpsace
	onto the columns of a DFT matrix. This provides further performance improvement in the resulting channel estimation/decontamination.
	

		The main contributions of this paper are listed as follows:		
		
		\begin{itemize}
			\item 
			We present in a concise and self-contained form the DL precoders and UL receivers from our previous works \cite{goettsch2021impact, kddi_uldl_precoding} based on instantaneous channel information obtained by UL pilot subspace projection, in contrast to large-scale fading decoding (LSFD)  \cite{8845768, demir2021cellfree} based on long term statistics. 
			The proposed schemes target directly the maximization of the {\em Optimistic Ergodic Rate} (OER) \cite{8304782}, while the motivation behind LSFD techniques is the maximization of the {\em Use-and-then-Forget} (UatF) lower bound on the ergodic rate \cite{Larsson-book}.
			
			\item We develop a new approximated UL-DL duality for the OER, and a consequent way to allocate power in the DL 
			to achieve OER very close to that of the UL, while reusing as DL precoders the same vectors computed for UL detection (such that no additional DL precoding computation is required).
			
			\item As said before, in order to enable UL pilot decontamination via subspace projection, 
			we propose a new method based on SRS pilots with Latin squares hopping sequences and R-PCA to estimate the dominant channel subspace for all connected UE-RU pairs (i.e., for all channels between any UE and all the RUs forming its user-centric cluster).
			
			\item Since a full system simulation including SRS pilots and R-PCA applied to each RU would be simply infeasible for 
			large networks with tens of RUs and hundreds of UEs, we propose a clever simulation technique based on evaluating by off-line Monte Carlo simulation the conditional 
			distribution of the estimated channel subspace at the output of the R-PCA conditioned on the pathloss coefficient 
			between the UE and the estimating RU. Then, large networks can be simulated by ``sampling'' from such conditional distribution in order to emulate the estimated channel subspace.  This enables the efficient simulation of large networks. 
			
			\item As a side result, we develop in Appendix \ref{OER-UatF} an achievable ergodic rate lower bound based on the insertion of 
			one dedicated pilot in the data slot. By comparing this lower bound to the OER we show that, 
			for the system parameters chosen in the present work  (which are very similar to several other published works on cell-free user-centric wireless networks), the OER provides indeed a close approximation of the actual achievable rate 
			while the popular UatF bound is over-conservative and may yield misleading conclusions. 
		\end{itemize}

	
	Our numerical results show that the proposed SRS hopping scheme and R-PCA subspace estimation method are able to closely approach  the performance of the pilot subspace projection with ideal subspace knowledge, which in turns has been shown to closely approach the performance of ideal partial CSI \cite{goettsch2021impact}. 
	This provides a first concrete and practical overall scheme for scalable operations in cell-free user-centric networks, including the estimation of long-term channel statistics, which is shown to essentially eliminate the problem of pilot contamination and (almost) achieve the performance of ideal localized channel state information.

	\section{System Model} \label{sec:ue_rrh_graph_and_ul_dl_model}
	
	We consider a cell-free wireless network with $K$ single-antenna UEs and $L$ RUs, each equipped with $M$ antennas. Both RUs and UEs are distributed randomly on a squared region on the two-dimensional plane. We assume OFDM modulation and that the communication channel follows the standard block-fading model (e.g., see \cite{9336188} and references therein) such that  the channel vector between any UE-RU pair is random but constant over time-frequency slots of $T=N_{\rm rb} \times N_{\rm sub}$ signal dimensions, where $N_{\rm rb}$ and $N_{\rm sub}$ denote the number of OFDM symbols in time and the number of OFDM subcarriers in frequency, respectively. 
	We can identify $T$ with the block length (in signal complex dimensions) of some adjacent resource block (RBs),  as defined by 5GNR  \cite{3gpp38211}, 
	used for UL and DL in TDD mode.  In particular, out of these $T$ signal dimensions, $\taudmrs$, $\tau_u$ and $\tau_d$ signal dimensions are used for UL 
	DMRS pilot, UL data and DL data transmission, respectively. 
	 Focusing on a given OFDM subcarrier, the $M \times 1$ channel vector for a given UE-RU pair is modeled as a correlated Gaussian vector \cite{9336188}, given by
		$
		\hv_{\ell,k} \sim \Cc \Nc \left( \zerov, \Sigma_{\ell,k} \right),
		$
		where $\zerov$ is the identically zero vector, and $\Sigma_{\ell,k} \in \CC^{M \times M}$ the covariance matrix. 
		For later use, we let $\beta_{\ell,k} = \frac{1}{M} \trace \left( \Sigma_{\ell, k} \right)$ denote the large-scale fading coefficient (LSFC),
		and $\Fm_{\ell,k}$ is the tall unitary matrix spanning the channel dominant subspace, i.e., its columns are given by the (unit-norm) eigenvectors of $\Sigma_{\ell, k}$ corresponding to the ``largest'' eigenvalues (this concept is clarified in 
		\cite{adhikary2013joint} and will be made precise in Sections \ref{sec:RPCA} and \ref{sec:simulations}, where we consider a particular correlation model).

	\subsection{Cluster formation and pilot assignment}
	
	We distinguish between SRS and DMRS pilots, and focus only on the allocation of the DMRS pilots in this section. These are used for per-slot (instantaneous) channel estimation 
	both in the UL and in the DL (via reciprocity).  The cluster formation and DMRS pilot assignment are carried out together.
	Let all UEs transmit with the same average energy per symbol $P^{\rm ue}$, and we define the system parameter 
	$\SNR \eqdef P^{\rm ue}/N_0,  \label{eq_snr}$
	where $N_0$ denotes the complex baseband noise power spectral density.
	There are a total of $\taudmrs$ orthogonal DMRS pilot sequences in the system, and we impose that 
	each RU assigns distinct pilots to all its associated UEs. Furthermore, the pilot of a given UE must be commonly assigned by all the RUs forming its serving 
	cluster \cite{9336188}.
	We let $\Uc_\ell$ denote the set of UEs associated to RU $\ell$ and $\Cc_k$ denote the set of RUs serving UE $k$.
	When a user $k$ wishes to join the system, through some beacon signal or some other location-based mechanism it selects its leading RU $\ell(k)$ as the one
	with the largest LSFC out of the set $\Lc_{\rm f}$ of RUs with free DMRS pilots, i.e.,
	$\ell(k) = \underset{\ell \in\Lc_{\rm f}}{\arg\max}\  \beta_{\ell,k}$,  provided that 
	\begin{gather}
		\beta_{\ell,k} \geq \frac{\eta}{M \SNR} , \label{eq:snr_threshold}
	\end{gather}
	where  $\eta > 0$ is a suitable threshold determining how much above the noise floor the useful signal in the presence of 
	maximum possible beamforming gain (equal to $M$) should be. 
	If such RU is not available, then the UE is declared in outage. Suppose that UE $k$ finds its leader RU $\ell(k)$ and it is allocated the DMRS pilot index $t_k \in [\taudmrs]$.\footnote{The set of integers from 1 to $n$ is denoted by $[n]$.} 
	Then, the cluster $\Cc_k$ is obtained sorting the RUs satisfying condition (\ref{eq:snr_threshold}) and having pilot $t_k$ still available in decreasing LSFC order, and adding them to the cluster until a maximum cluster size $Q$ is reached, where $Q$ is a design parameter imposed to limit the computational complexity of each cluster processor. 
	As a result, for all UEs $k$ not in outage, $1 \leq  |\Cc_k| \leq Q$ and for all RUs $\ell$ we have $|\Uc_\ell| \leq \taudmrs$. 
	
	\begin{figure}[ht]
		\centerline{\includegraphics[trim={140 130 100 82}, clip,width=6cm]{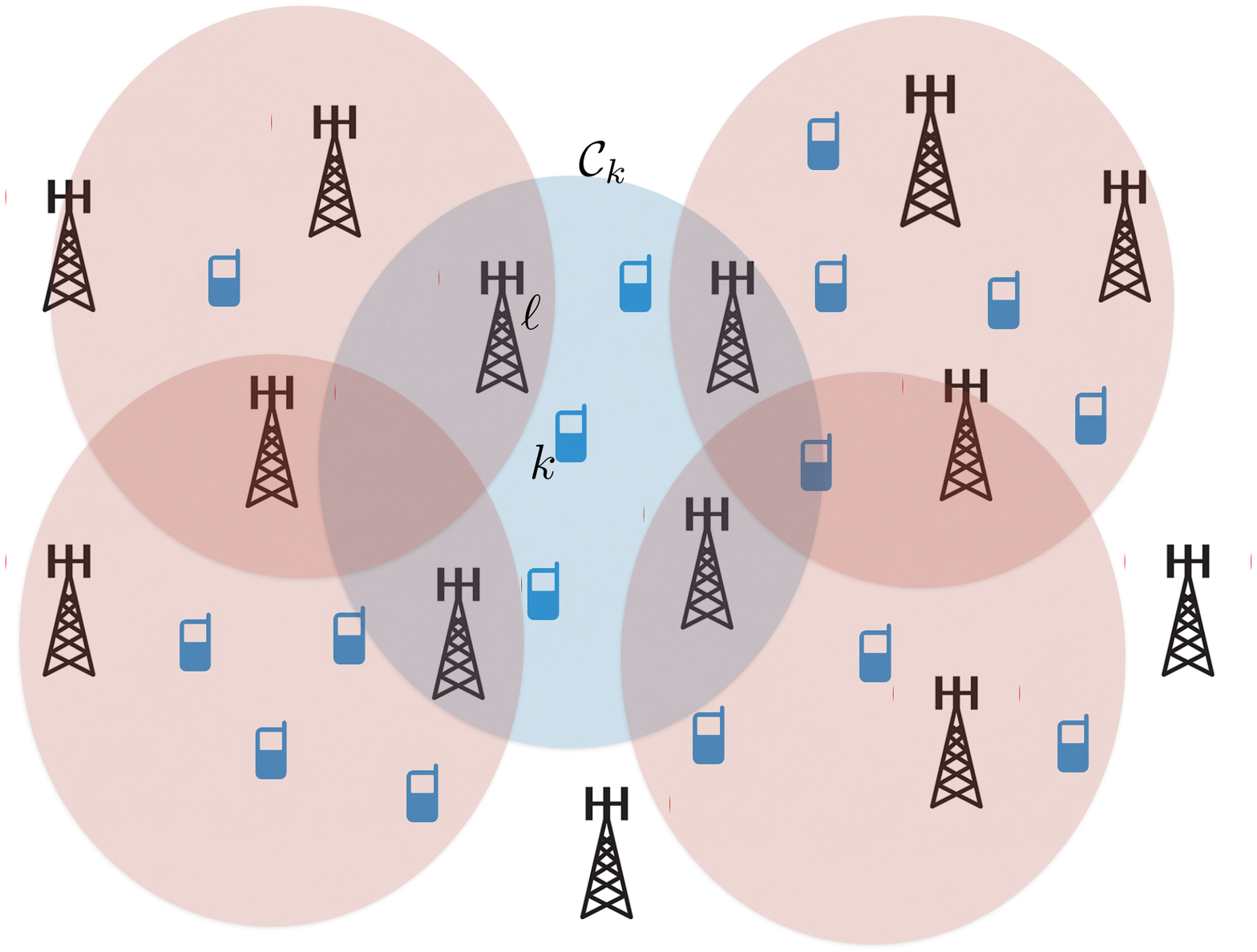} \hspace{.3cm} \includegraphics[trim={100 178 140 178}, clip, width=9cm]{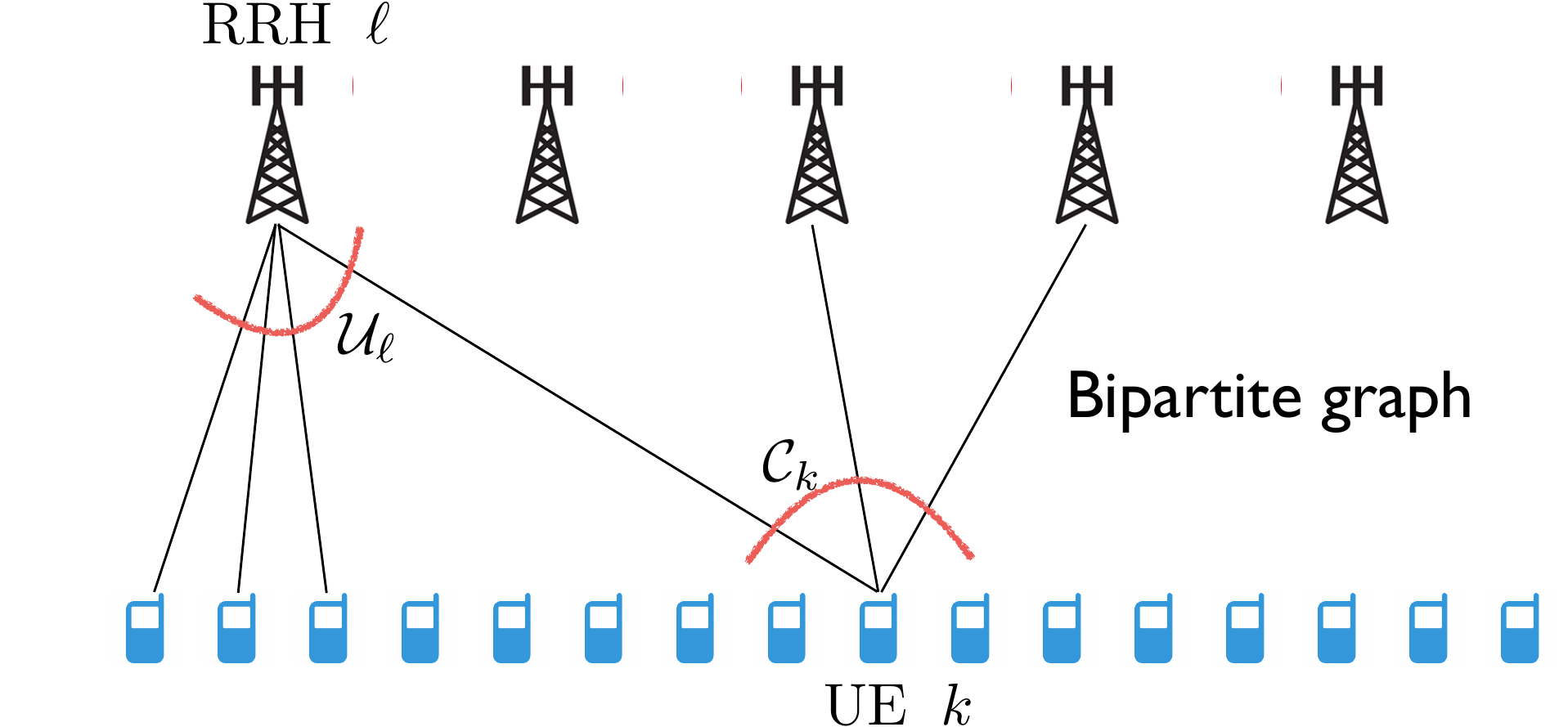}}
		\vspace{-.0cm}
		\caption{An example of user-centric clusters and UE-RU association. The sets $\Uc_\ell$ of UEs associated to each RU $\ell$ and the sets $\Cc_k$ of RU serving UE $k$ define a bipartite graph with RU and UE vertices and edges $(k,\ell)$ for all $k \in [K]$ and $\ell \in [L]$ such that 
			$k \in \Uc_\ell$ and $\ell \in \Cc_k$.}
		\vspace{-.0cm}		
		\label{clusters}
	\end{figure}
	
	We can visualize the resulting UE-RU association as a bipartite graph $\Gc$ with two classes of nodes (UEs and RUs) such that the neighborhood of UE nodes $k$ is $\Cc_k  \subseteq [L]$ and the neighborhood of the RU nodes $\ell$ is $\Uc_\ell \subseteq [K]$ (see Fig.~\ref{clusters}). 
	The set of edges of $\Gc$ is denoted by $\Ec$. We let $\HH \in \CC^{LM \times K}$ denote the overall channel matrix between all UEs and all the RU antennas. 
	This matrix is formed by $M \times 1$ blocks $\hv_{\ell,k}$, denoting the channel between the $M$ antennas of RU $\ell$ and the single antenna of UE $k$. 
	 Due to the local cluster knowledge about the DMRS pilot allocation, 
		each RU $\ell$ can only estimate the channel vectors of the users in $\Uc_\ell$, i.e., 
		the known channel vectors (up to estimation noise) are the vectors $\hv_{\ell,k}$ such that $(\ell,k) \in \Ec$.
	Therefore,  each RU has a partial (noisy) view of the matrix $\HH$.  We define $\Uc(\Cc_k) = \bigcup_{\ell \in \Cc_k} \Uc_\ell$
	as the set of users served by at least one RU in $\Cc_k$.  Notice that, by construction, $k \in \Uc(\Cc_k)$. 
	We define the part of the channel matrix $\HH$ known at the processor for cluster $\Cc_k$ as  $\HH(\Cc_k)$, 
	of  the same dimensions of $\HH$,  such that the $(\ell, j)$ block of dimension $M \times 1$ of $\HH(\Cc_k)$ is equal to $\hv_{\ell,j}$ for all  
	$\ell \in \Cc_k$  and $j \in \Uc_\ell$, and to $\zerov$ otherwise.
	Notice that $\HH(\Cc_k)$ has $|\Uc(\Cc_k)|$ non-all-zero columns.
	
	\begin{example}
		Consider the simple case of $L = 2$ and $K = 6$.
		Let's focus on user $k = 3$, for which $\Cc_3 = \{1,2\}$. We have $\Uc_1 = \{1,2,3,4\}$ and $\Uc_2 = \{3,4,5,6\}$, therefore $\Uc(\Cc_3) = \{1,2,3,4,5,6\}$. 
		The complete channel matrix is given by 
		\[ \HH = \left [ \begin{array}{cccccc}
			\hv_{1,1} & \hv_{1,2} & \hv_{1,3} & \hv_{1,4} & \hv_{1,5} & \hv_{1,6} \\
			\hv_{2,1} & \hv_{2,2} & \hv_{2,3} & \hv_{2,4} & \hv_{2,5} & \hv_{2,6} \end{array} \right ], \]
		while the {\em partial} cluster-centric channel matrix $\HH(\Cc_3)$ is given by 
		\[ \HH(\Cc_3) = \left [ \begin{array}{cccccc}
			\hv_{1,1} & \hv_{1,2} & \hv_{1,3} & \hv_{1,4} & \zerov & \zerov \\
			\zerov & \zerov & \hv_{2,3} & \hv_{2,4} & \hv_{2,5} & \hv_{2,6} \end{array} \right ]. \]
		\hfill $\lozenge$
	\end{example}
	

	\subsection{Uplink data transmission} 
	
	We let $\sss^{\rm ul} \in \CC^{K \times 1}$  denote the vector of modulation symbols transmitted collectively by the $K$ users over the UL at a given symbol time (channel use).
	The symbols $s_k^{\rm ul}$ ($k$-th components of $\sss^{\rm ul}$) are i.i.d. with mean zero and unit variance. 
	The observation (samples) at the $LM$ RUs antennas for a single UL channel use is given by 
	\begin{equation} 
		\yy^{\rm ul} = \sqrt{\SNR} \; \HH \sss^{\rm ul}   + \zz^{\rm ul}, \label{ULchannel}
	\end{equation}
	where $\zz^{\rm ul}$ has i.i.d. components $\sim \Cc\Nc(0,1)$. 
	The goal of the cluster $\Cc_k$ processor  is to produce an estimate of 
	$s^{\rm ul}_k$ which is treated by the channel decoder for user $k$ as the soft-output of a virtual single-user channel.  
	As in most works on (cell-free) massive MIMO, we restrict the cluster processing to be linear and 
	define the receiver unit-norm vector $\vvv_k \in \CC^{LM \times 1}$ formed by blocks
	$\vv_{\ell,k}  \in \CC^{M \times 1}, \ \ell \in [L]$, such that $\vv_{\ell,k} = \zerov$ if $(\ell,k) \notin \Ec$. 
	The non-zero blocks $\vv_{\ell,k} :  \ell \in \Cc_k$ can be suitably defined according to some combining scheme (to be detailed later).
	The corresponding scalar combined observation for symbol $s^{\rm ul}_k$ is given by 
	\begin{eqnarray}
		r^{\rm ul}_k  & = & \vvv_k^\herm \yy^{\rm ul} 
		= \sqrt{\SNR} \vvv_k^\herm \hh_k s^{\rm ul}_k   + \sqrt{\SNR} \vvv_k^\herm \HH_k \sss_k^{\rm ul}  + \vvv_k^\herm \zz^{\rm ul} ,  \label{received-UL}
	\end{eqnarray}
	where we let $\hh_k$ denote the $k$-th column of $\HH$, where $\HH_k$ is the matrix $\HH$ after elimination of the $k$-th column, 
	and $\sss_k^{\rm ul} $ is the vector $\sss^{\rm ul}$ after elimination of the $k$-th element. 
	
	The resulting UL {\em instantaneous} SINR is given by 
	\begin{eqnarray} 
		\SINR^{\rm ul}_k  & = & \frac{  |\vvv_k^\herm \hh_k|^2 }{ \SNR^{-1}  + \sum_{j \neq k} |\vvv_k^\herm \hh_j |^2 }.  \label{UL-SINR-unitnorm}
	\end{eqnarray}
	As a performance measure, in this paper we consider the OER given by \cite{8304782}
	\begin{eqnarray}
		R^{\rm ul}_k = \EE [ \log (1 + \SINR^{\rm ul}_k) ], \label{ergodic_rate_ul}
	\end{eqnarray}
	where the expectation is with respect to the small-scale fading, while conditioning on the long-term 
	variables (LSFCs, placement of UEs and RUs on the plane, and cluster formation).
This rate metric is achievable if the receiver knows the phase of the useful signal term $\vvv_k^\herm \hh_k$ and the 
		value of $\SINR^{\rm ul}_k$ on each time-frequency slot. 
		However, there is no ``converse'' result that says that this is {\em not achievable}  
		under some relaxed channel state knowledge requirement. In particular, {\em block non-coherent schemes} \cite{hanzo2011mimo}
		or {\em universal decoding scheme} \cite{feder1998universal} may closely approach the OER. 
		As a matter of fact, the widely used UatF  lower bound on the ergodic rate
		\cite{Larsson-book,9336188} may be overly pessimistic when the
		useful signal coefficient  does not show enough ``hardening'', i.e., when the squared mean
		$|\EE[\vvv_k^\herm \hh_j]|^2$ is not much larger than the variance Var$(\vvv_k^\herm \hh_k)$
		(e.g., see \cite{8304782}). In Appendix \ref{OER-UatF} we develop a simple
		achievable rate lower bound based on conditioning with respect to a pilot symbol
		included in the data payload, and show that, for the system parameters considered {\BLUE in} this paper, 
		this lower bound is indeed very close to the OER, while the UatF is considerably worse. 
		Obviously, the best possible achievable rate is trapped between
		the OER and the lower bound of Appendix \ref{OER-UatF}, and therefore is very well approximated by 
		the OER in our case. The reason why we use the OER rather than the bound of  Appendix \ref{OER-UatF} is that the OER is significantly simpler 
		to evaluate (by Monte Carlo simulation) and also because, as we shall see in Section \ref{UL-DL-duality}, for the OER we can prove  
		an approximated UL-DL duality, which significantly simplifies the design of the DL precoding schemes.

	\subsection{Downlink data transmission}
	
	In this case, one DL channel use
	at the receiver of UE $k$ yields the scalar signal sample
	\begin{equation} 
		y_k^{\rm dl} = \hh_k^\herm \xx  + z_k^{\rm dl},  \label{DLchannel}
	\end{equation}
	where the transmitted vector $\xx \in \CC^{LM \times 1}$ is formed by all the signal samples sent collectively from the RUs. 
	Without loss of generality, we let the noise sample at each $k$-th UE receiver be $z_k^{\rm dl} \sim \Cc\Nc(0, \SNR^{-1})$. 
	Let $\sss^{\rm dl} \in \CC^{K \times 1}$ denote the vector of modulation symbols for the $K$ users 
	(independent with mean zero and variance $q_k \geq 0$).  Under a general DL linear precoding scheme, we have
	$\xx = \UU \sss^{\rm dl}$,  where $\UU \in \CC^{LM \times K}$ is the overall DL precoding matrix with unit-norm columns. 
	In order to preserve the per-cluster processing condition, this is a block matrix formed by $M \times 1$ blocks $\uv_{\ell,k}$ such that 
	$\uv_{\ell,k} = \zerov$ if $(\ell, k) \notin \Ec$.  The non-zero blocks $\uv_{\ell,k} :  (\ell,k) \in \Ec$ can be suitably defined (to be detailed later).  
	It follows that \eqref{DLchannel} can be written as
	\begin{eqnarray}
		y^{\rm dl}_k & = & \hh_k^\herm \uu_k s^{\rm dl}_k   + \sum_{j \neq k} \hh_k^\herm \uu_j s^{\rm dl}_j  + z^{\rm dl}_k, 
	\end{eqnarray}
	where $\uu_k$ is the $k$-th column of $\UU$ with corresponding  DL SINR
	\begin{eqnarray}
		\SINR^{\rm dl}_k & = & \frac{|\hh_k^\herm \uu_k|^2 q_k}{\SNR^{-1} + \sum_{j\neq k}   |\hh_k^\herm \uu_j|^2 q_j }. \label{DL-SINR} 
	\end{eqnarray}
	As in the UL, we consider the DL optimistic ergodic rate
	\begin{eqnarray}
		R^{\rm dl}_k = \EE [ \log (1 + \SINR^{\rm dl}_k) ]. \label{ergodic_rate_dl}
	\end{eqnarray}
	Using the fact that $\UU$ has unit-norm columns, we find 
	$\trace \left ( \EE [ \xx \xx^\herm ] \right )  = 
	\sum_{k=1}^K q_k$.
	Imposing that the total Tx powers in the UL and DL are both equal to $K P^{\rm ue}$, with the above normalizations 
	we find  the condition $\sum_{k=1}^K q_k = K$ for the individual users' DL data stream power coefficients. 
	
	\section{UL Detection and DL Precoding Schemes}  \label{sec:schemes}
	
	Our goal is to provide extensive system performance comparisons under several alternatives for UL detection (choice of the $\vvv_k$'s) and 
	DL precoding (choice of the $\uu_k$'s) linear schemes with actual CSI information obtained from the UL DMRS pilots.
	For the sake of clarity, in this section we shall illustrate the considered detection and precoding schemes assuming {\em ideal partial CSI}, as explained in sections \ref{sec:intro} and \ref{sec:ue_rrh_graph_and_ul_dl_model}.
	Then, in Section \ref{sec:pilot} we focus on the actual CSI estimation scheme. The results of Section \ref{sec:simulations} are then obtained by replacing 
	the {\em actual estimated} CSI into the {\em ideal partial} CSI expressions of this section. 
	
	\subsection{UL detection schemes}
	
	
	{\bf Cluster-level Zero-Forcing (CLZF).}
	Let $\overline{\HH}(\Cc_k) \in  \CC^{|\Cc_k|M \times |\Uc(\Cc_k)|}$ denote the channel matrix obtained from $\HH(\Cc_k)$ after 
	removing all columns $j \neq \Uc(\Cc_k)$ and all blocks of $M$ rows corresponding to RUs $\ell \notin \Cc_k$.  
	Let $\overline{\hh}_k(\Cc_k)$ and $\overline{\HH}_k(\Cc_k)$ denote the column of $\overline{\HH}(\Cc_k)$ corresponding to user $k$
	and the residual matrix after extracting such column from $\overline{\HH}(\Cc_k)$, respectively. 
	Consider the  singular value decomposition (SVD)  $\overline{\HH}_k(\Cc_k) = \overline{\AA}_k \overline{\SSS}_k \overline{\BB}_k^\herm$, 
	where the columns of the tall unitary matrix $\overline{\AA}_k$ form an orthonormal basis for the column span of 
	$\overline{\HH}_k(\Cc_k)$, i.e., the subspace spanned by the known channels of the users interfering with user $k$ at cluster $\Cc_k$. 
	The orthogonal projector onto the orthogonal complement of such {\em interference subspace} is given by  
	$\overline{\PP}_k = \Id - \overline{\AA}_k \overline{\AA}_k^\herm$. Define the  unit-norm vector 
	$\overline{\vvv}_k = \overline{\PP}_k \overline{\hh}_k(\Cc_k) / \| \overline{\PP}_k \overline{\hh}_k(\Cc_k) \|$. 
	The CLZF receiver vector $\vvv_k$ is given by expanding $\overline{\vvv}_k$ by reintroducing the missing 
	blocks of all-zero $M \times 1$ vectors $\zerov$ in correspondence of the RUs $\ell \notin \Cc_k$. 
	
	Due to the antenna correlation introduced by the scattering model in the simulations section, it may happen that
		some UEs  $k' \in \Uc_\ell, k' \neq k$ have channels $\hh_{k'}(\Cc_k)$ co-linear with $\hh_k(\Cc_k)$ with probability 1 (with respect to the small-scale fading realizations).  
	In this case, the CLZF would yield  $\vvv_k^\herm \hh_k(\Cc_k) = 0$. 
	In order to avoid this ``zero-forcing outage'', 
	we employ a simple scheme: if the cluster $\Cc_{k}$ detects a UE $k' \in \Uc(\Cc_{k})$ with a co-linear channel, it computes the CLZF combining vector excluding $\hh_{k'}(\Cc_k)$, i.e., it removes the row corresponding to UE $k'$ from the matrix $\overline{\HH}_k(\Cc_k)$. 
	This scheme allows interference from UE $k'$ (as it would happen with simple maximum ratio combining), 
	but still eliminates interference from the UEs in $\Uc(\Cc_{k}) \setminus \{k, k'\}$. 
	Note that co-linearity of $\hh_{k}(\Cc_k)$ and $\hh_{k'}(\Cc_k)$ does not imply in general that $\hh_{k}(\Cc_{k'})$ and $\hh_{k'}(\Cc_{k'})$ from the perspective of cluster $\Cc_{k'}$ are co-linear, as $\Cc_k$ and $\Cc_{k'}$ may contain different sets of RUs.

	{\bf Local LMMSE with cluster-level combining.} 
	In this case, each RU $\ell$ makes use of locally computed receiving vectors $\vv_{\ell,k}$ for 
	its users $k \in \Uc_\ell$. 
	Letting 
	\begin{equation}
		\yv_\ell^{\rm ul} =  \sqrt{\SNR} \sum_{j=1}^K \hv_{\ell,j} s_j^{\rm ul}   + \zv_\ell^{\rm ul},  
		\label{receivedsignal-ell}
	\end{equation}
	the $M \times 1$ block of $\yy^{\rm ul}$ corresponding to RU $\ell$,  RU $\ell$ computes the local detector
	$r^{\rm ul}_{\ell,k} = \vv_{\ell,k}^\herm \yv_\ell^{\rm ul}$ for each $k \in \Uc_\ell$ and sends
	the symbols $\{r^{\rm ul}_{\ell,k} : k \in \Uc_\ell\}$ to the DUs implementing cluster-based combining. Then, the processor for cluster $\Cc_k$ (implemented at some DU)
	computes the cluster-level combined symbol 
	\begin{equation} 
		r^{\rm ul}_k = \sum_{\ell \in \Cc_k} w^*_{\ell,k} r^{\rm ul}_{\ell,k} = \wv^\herm_k \rv^{\rm ul}_k,   \label{combining_perfect_csi}
	\end{equation}
	where we define $\rv^{\rm ul}_k = \{ r^{\rm ul}_{\ell,k} : \ell \in \Cc_k\}$ and where $\wv_k$ is a vector of combining coefficients. 
	In this paper we consider the case where the $\vv_{\ell,k}$'s are the linear MMSE (LMMSE) receiver vectors given the partial local CSI at RU $\ell$, while treating the out of cluster interference component (whose CSI is unknown to RU $\ell$) as an increase of the variance of the white Gaussian noise term, 
		given by 
		$
		\sigma^2_\ell =  1 + \SNR \sum_{j \neq \Uc_\ell}  \beta_{\ell,j} \label{sigmaell} 
		$, see details in  \cite{goettsch2021impact}.
	Under this assumption, the LMMSE receiving vector for user $k$ at RU $\ell$ is given by 
	\begin{equation} 
		\vv_{\ell,k} = \left ( \sigma_\ell^2 \Id + \SNR \sum_{j \in \Uc_\ell} \hv_{\ell,j} \hv_{\ell,j}^\herm \right )^{-1} \hv_{\ell,k}.  \label{eq:lmmse}
	\end{equation}
	It follows that the aggregate channel observation at the $k$-th cluster processor after local LMMSE combining can be written in the form
	\begin{equation}
		\rv^{\rm ul}_k = \sqrt{\SNR}  \left ( \av_k s^{\rm ul}_k  +  \Gm_k \sv_k^{\rm ul} \right ) + \zetav_k 
	\end{equation} 
	where we define $g_{\ell,k,j} = \vv_{\ell,k}^\herm \hv_{\ell,j}$, the vector $\av_k = \{ g_{\ell, k,k} : \ell \in \Cc_k\}$,  
	the matrix $\Gm_k$ of dimension $|\Cc_k| \times ( | \Uc(\Cc_k)| - 1)$ containing 
	elements $g_{\ell, k, j}$ in position corresponding to  RU $\ell$ and UE $j$ (after a suitable index reordering) 
	if $(\ell,j) \in \Ec$, and zero elsewhere, the $( |\Uc(\Cc_k)| - 1) \times 1$ vector $\sv_k^{\rm ul}$ of the 
	symbols of all UEs $j \in \Uc(\Cc_k) : j \neq k$, and   
		the noise plus out-of-cluster interference vector $ \zetav_k = \left\{ \vv_{\ell,k}^\herm \xiv_\ell : \ell \in \Cc_k \right\}$ with covariance matrix $\Dm_k = \diag \left \{ \sigma_\ell^2 \|\vv_{\ell,k}\|^2 : \ell \in \Cc_k \right \}$, and $\xiv_\ell$ assumed  $\sim \Cc \Nc \left( \zerov, \sigma_\ell^2 \Id \right)$. 
	
	The corresponding SINR for UE $k$ with combining (\ref{combining_perfect_csi}) is given by 
	\begin{equation} 
		\SINR^{\rm ul-nom}_k = \frac{\SNR \; \wv_k^\herm \av_k \av_k^\herm \wv_k}{\wv_k^\herm \Gammam_k  \wv_k},  \label{SINRnom_perfect_csi}
	\end{equation}
	where  we define $\Gammam_k = \Dm_k  + \SNR \; \Gm_k \Gm^\herm_k$. 
	The maximization of this {\em nominal SINR}~\footnote{This is called ``nominal'' since it is the SINR under the assumption that the out-of-cluster interference 
		is replaced by a corresponding white noise term with the same per-component variance.}  
	with respect to the vector of combining coefficients $\wv_k$ is a classical generalized Rayleigh quotient maximization problem, solved by 
	finding the generalized eigenvector of the maximum generalized eigenvalue of the matrix pencil 
	$(\av_k \av_k^\herm, \Gammam_k )$.  Since the matrix $\av_k \av_k^\herm$ has rank 1 and therefore it has only one non-zero eigenvalue, 
	it is immediate to see that the solution is given by  $\wv_k = \Gammam_k^{-1} \av_k$. Notice that the {\em actual} SINR resulting from this scheme 
	is again given by (\ref{UL-SINR-unitnorm}) where $\vvv_k$ is obtained by stacking the components $w_{\ell,k} \vv_{\ell,k}$ into an $LM \times 1$ vector (with the zero blocks
	corresponding to $\ell \neq \Cc_k$) and normalizing the whole vector to have unit norm.

	\begin{remark}  {\bf Connection to LSFD.} 
		It is worthwhile to point out the difference between the local LMMSE with cluster-level combining 
		considered in this paper and the  LSFD approach proposed in \cite{8845768} and \cite{demir2021cellfree}. 
		Both schemes are based on local (per-RU) linear detection and cluster-level combining. 
		The difference is in the computation of the combining coefficients. In LSFD, the goal is to maximize the
		UatF bound  which depends only on the second-order statistics of the channel coefficients after local combining. 
		Therefore, in our notation, the LSFD coefficients are given by  $\wv^\text{LSFD}_k = (\Gammam_k^\text{LSFD})^{-1} \EE[\av_k]$,
		where   $\Gammam_k^\text{LSFD} = \Dm_k  + \SNR \; \EE[ \Gm_k \Gm_k^\herm]$. 
		Notice that LSFD requires the estimation of the expected quantities (e.g., by some sliding window local averaging of the instantaneous
		values of $\av_k$ and $\Gm_k \Gm_k^\herm$). 
		This is because the expected quantities  
		are generally unknown a priori (despite claims in  \cite{8845768,demir2021cellfree} based on the unrealistic assumption of known second-order statistics). 
		In contrast, our scheme directly makes use of such per-slot instantaneous 
		values that can be sent on each slot from RUs to the cluster processors. We will provide a comparison of the LSFD and the proposed schemes in the results section.
		\hfill $\lozenge$ \label{remark_lsfd}
	\end{remark}
	
	\vspace{-5mm}

	\subsection{DL precoding and power allocation}     \label{UL-DL-duality}

	We propose to reuse as DL precoders the same vectors calculated for UL detection, i.e., $\uu_k = \vvv_k$ for all $k \in [K]$. 
	This approach greatly reduces the computation complexity of the whole system, since a single vector per user
	for both UL and DL must be computed. The motivation of this choice relies on the following UL/DL ``nominal SINR'' duality result.
	
	{\bf Uplink-downlink nominal SINR duality for partially known channels.}
	Under the ideal partial CSI assumption, only the channel vectors $\hv_{\ell,k}$ with $(\ell,k) \in \Ec$ are (perfectly) known, while for all 
	$(\ell,k) \notin \Ec$ the channels are completely unknown. 
	%
	Considering the UL SINR given in (\ref{UL-SINR-unitnorm}), we notice that the term at the numerator
	\begin{equation} 
		\theta_{k,k} = \left | \vvv_k^\herm \hh_k \right |^2 = \left | \sum_{\ell\in \Cc_k} \vv_{\ell,k}^\herm \hv_{\ell,k} \right |^2 \label{thetakk}
	\end{equation}
	contains only known channels and therefore is fully known by the system.  The terms at the denominator take on the form
	\begin{eqnarray}
		\theta_{j,k} & = & \left | \vvv_k^\herm \hh_j \right |^2 
		= \left | \sum_{\ell \in \Cc_k \cap \Cc_j} \vv_{\ell,k}^\herm \hv_{\ell,j}  + \sum_{\ell \in \Cc_k \setminus \Cc_j} \vv_{\ell,k}^\herm \hv_{\ell,j} \right |^2 \label{known-unknown}
	\end{eqnarray}
	where for $\ell \in \Cc_k \cap \Cc_j$ the channel $\hv_{\ell,j}$ is known, while for $\ell \in \Cc_k \setminus \Cc_j$ the channel $\hv_{\ell,j}$ is not known. 
	Taking the conditional expectation of the term in (\ref{known-unknown}) given the known channel state information
	at the cluster processor for $\Cc_k$  (denoted for brevity by $\Ec_k$), we find
	\begin{equation} 
			\EE[ \theta_{j,k} | \Ec_k] = \left | \sum_{\ell \in \Cc_k \cap \Cc_j} \vv_{\ell,k}^\herm \hv_{\ell,j} \right |^2 + 
			\sum_{\ell \in \Cc_k\setminus \Cc_j} \vv_{\ell,k}^\herm \Sigma_{\ell,j} \vv_{\ell,k}.  
	\end{equation} 
	Finally, under the isotropic assumption, we replace the actual covariance matrix of the unknown channels with
	a scaled identity matrix with the same trace, i.e., 
	$\Sigma_{\ell,j} \leftarrow \beta_{\ell, j} \Id$, and 
	using the fact that $\vvv_k$ is a unit-norm vector and that (approximately) the  $M \times 1$ blocks $\vv_{\ell,k}$ have the same norm, 
	we can further approximate
	$\| \vv_{\ell,k}\|^2 \approx \frac{1}{|\Cc_k|}$. 
	Therefore, we define the interference coefficients $\widetilde{\theta}_{j,k}$ such that
	\begin{equation} 
		\EE[ \theta_{j,k} | \Ec_k] \approx \widetilde{\theta}_{j,k} \eqdef 
		\left | \sum_{\ell \in \Cc_k \cap \Cc_j} \vv_{\ell,k}^\herm \hv_{\ell,j} \right |^2 + \frac{1}{|\Cc_k|} \sum_{\ell \in \Cc_k\setminus \Cc_j} \beta_{\ell,j}.  \label{new-theta-jk}
		%
	\end{equation}
	and the resulting  {\em nominal} UL SINR  as
	\begin{eqnarray} 
		\SINR_k^{\rm ul-nom} 
	& \eqdef & \frac{\theta_{k,k}}{\SNR^{-1} + \sum_{j\neq k} \widetilde{\theta}_{j,k} } \label{eq:nom-ul-SINR}
\end{eqnarray}
Next, we consider the DL SINR given in (\ref{DL-SINR}) under the assumption $\uu_k = \vvv_k$ for all $k \in [K]$. 
The numerator takes on the form $\theta_{k,k} q_k$, where $\theta_{k,k}$ is again given by \eqref{thetakk}.
Focusing on the terms at the denominator, after a calculation analogous to what done before and 
taking into account that the transmit vectors $\uu_j = \vvv_j$ are non-zero only for the $M \times 1$ blocks 
corresponding to RUs $\ell \in \Cc_j$ (i.e., for the RUs of the cluster serving UE $j$), we eventually obtain the expression of a nominal DL SINR in the form
\begin{eqnarray} 
	\SINR_k^{\rm dl-nom} 
	& \eqdef & \frac{\theta_{k,k} q_k}{\SNR^{-1} + \sum_{j\neq k} \widetilde{\theta}_{k,j} q_j} ,
\end{eqnarray}
where $\widetilde{\theta}_{k,j}$ takes on the form \eqref{new-theta-jk} after exchanging the indices $k$ and $j$.  
Given the symmetry of the coefficients, a UL-DL duality exists for  the nominal SINRs. 
In particular, the DL Tx power allocation coefficients $\{q_k : k \in [K]\}$ can be computed such that the per-user 
DL nominal SINRs coincide with the corresponding UL nominal SINRs. 
Explicitly, define for notation convenience  $\gamma_k := \SINR^{\rm ul-nom}_k$ where the latter is given by \eqref{eq:nom-ul-SINR}. 
The system of equations in the power allocation vector $\qv = \{q_k\}$ given by $\SINR^{\rm dl-nom}_k = \gamma_k, \; k \in [K]$
can be rewritten in the convenient linear system form as \cite{Viswanath-Tse-TIT03}
\begin{equation} 
	\left ( \Id  - \diag(\muv) \Thetam \right ) \qv  = \frac{1}{\SNR} \muv,  \label{eq_q_duality}
\end{equation} 
where $\muv$ is the $K\times 1$ vector with elements  $\mu_k = \gamma_k/((1 + \gamma_k) \theta_{k,k})$
and the matrix $\Thetam$ has $(k,j)$ elements $\theta_{k,k}$ on the diagonal and $\widetilde{\theta}_{k,j}$ for $j \neq k$ off the diagonal. 
It can be shown \cite{Viswanath-Tse-TIT03} that the solution $\qv^\star$ of (\ref{eq_q_duality})
has non-negative components.  It is also immediate to show that the solution satisfies $\sum_{k=1}^K q^\star_k = K$, i.e., the total transmit power in UL and DL 
is the same.  Based on this duality,  letting $\uu_k = \vvv_k$ and using the DL power allocation $\qv^\star$, the same nominal SINR in UL and DL can be achieved. 
While this does not ensure exact duality of the instantaneous SINRs, we shall see that the actual optimistic ergodic rates achieved
in UL and DL with this criterion are very similar for all the considered precoding/detection schemes.  
\section{UL Pilot Transmission and Channel Estimation}  \label{sec:pilot}

In practice, ideal partial CSI is not available and the channels $\{\hv_{\ell,k} : (\ell,k) \in \Ec\}$ must be estimated from UL pilots. 
We assume that $\taudmrs$ signal dimensions per slot are dedicated to UL pilots and define a codebook of $\taudmrs$ orthogonal pilot 
sequences.  The pilot field received at RU $\ell$ is given by the $M \times \taudmrs$ matrix 
\begin{equation} 
	\Ym_\ell^{\rm pilot} = \sum_{i=1}^K \hv_{\ell,i} \left(\phiv_{t_i}^{ \text{DMRS} }\right)^\herm + \Zm_\ell^{\rm pilot} \label{Y_pilot}
\end{equation}
where $\Zm_\ell^{\rm DMRS}$ is AWGN with elements i.i.d. $\sim \Cc\Nc(0, 1)$, and $\phiv_{t_i}^{ \text{DMRS}}$ denotes the pilot  vector of dimension $\taudmrs$ used by UE $i$ at the current slot, normalized such that
$\| \phiv_{t_i} \|^2 = \taudmrs\SNR$ for all $t_i \in [\taudmrs]$. The standard {\em Least-Squares} channel estimation used in overly many papers on massive MIMO
consists of ``pilot matching'', i.e., RU $\ell$ produces the channel estimates \begin{eqnarray} 
	\widehat{\hv}^{\rm pm}_{\ell,k} & = & \frac{1}{\taudmrs \SNR} \Ym^{\rm pilot}_\ell \phiv_{t_k}  
	= \hv_{\ell,k}  + \sum_{ i \neq k : t_i = t_k} \hv_{\ell,i}  + \widetilde{\zv}_{t_k,\ell}   \label{chest}
\end{eqnarray} 
by right-multiplication of the pilot field by $\phiv_{t_k}$  for all $k \in \Uc_\ell$,  where $\widetilde{\zv}_{t_k,\ell}$ is $M \times 1$ Gaussian i.i.d. with components $\Cc\Nc(0, \frac{1}{\taudmrs\SNR})$ and where  $\sum_{i \neq k : t_i = t_k} \hv_{\ell,i}$ is the pilot contamination term, i.e., the contribution of the channels from users $i \neq k$ using the same pilot $t_k$ (co-pilot users).  Assuming that the subspace information $\Fm_{\ell,k}$ of all $k \in \Uc_\ell$ is known, we consider also the ``subspace projection'' (SP) 
pilot decontamination scheme given by the orthogonal projection of $\widehat{\hv}^{\rm pm}_{\ell,k}$ onto the subspace spanned by the columns of 
$\Fm_{\ell,k}$, i.e., 
\begin{eqnarray}
	\widehat{\hv}^{\rm sp}_{\ell,k} & = & \Fm_{\ell,k}\Fm_{\ell,k}^\herm \widehat{\hv}^{\rm pm}_{\ell,k} =  \hv_{\ell,k}  + \Fm_{\ell,k}\Fm_{\ell,k}^\herm \left ( \sum_{ i \neq k : t_i = t_k} \hv_{\ell,i} \right ) + \Fm_{\ell,k} \Fm_{\ell,k}^\herm \widetilde{\zv}_{t_k,\ell}.   \label{chest1}
\end{eqnarray}
Writing explicitly the pilot contamination term after the subspace projection, it is immediate to show that its covariance matrix is given by 
	$
	\Sigmam_{\ell,k}^{\rm co}  = \sum_{i \neq k : t_i = t_k} \Fm_{\ell,k} \Fm_{\ell,k}^\herm \Sigma_{\ell,i} \Fm_{\ell,k} \Fm^\herm_{\ell,k} .
	$
	When $\Fm_{\ell,k}$ and $\Fm_{\ell,i}$ are nearly mutually orthogonal, i.e. $\Fm_{\ell,k}^\herm \Fm_{\ell.i} \approx \zerov$,
	the subspace projection is able to significantly reduce the pilot contamination effect.

\section{Subspace Estimation with UL SRS pilots} \label{sec:srs_rpca}

In order to implement the DMRS channel estimation with pilot decontamination based on channel subspace projection in (\ref{chest1}), 
the dominant subspace of each user channel $\hv_{\ell,k}$ with $k \in \Uc_\ell$ must be estimated at each RU $\ell$. 
Such subspaces are long-term statistical properties 
	that are frequency-independent, and
	depend on the geometry of the propagation, which is assumed to remain essentially invariant over sequences of many consecutive slots. 
	Further, assuming ULAs or UPAs and owing to the Toeplitz or block-Toeplitz structure of the channel covariance matrices, 
for large $M$ such subspaces are nearly spanned by subsets of the DFT columns (see \cite{adhikary2013joint}). 

In this section, we propose to use SRS pilots for channel subspace estimation, 
where we assume that each user sends just {\em one} SRS pilot symbol 
per slot, hopping over the subcarriers in multiple slots. In this sense, ``orthogonal SRS pilots'' indicate symbols sent by different users on different subcarriers in the same slot. 
We reserve a grid of $N$ distinct subcarriers for the SRS pilot in each slot, such that in any slot time there exist 
$N$ orthogonal SRS pilots.  We say that two users have a SRS pilot collision at some slot, when they transmit their SRS pilot symbol on the same subcarrier. 
The allocation of SRS pilot sequences to users is done in order to minimize the pilot collisions between users at the RUs forming their clusters. 
It may happen that some user $k$ collides on some slot $s$ with some user $j$, and that user $j$ is much closer to  RU $\ell \in \Cc_k$ than user $k$. In this case, the measurement at RU $\ell$ relative to user $k$ is heavily interfered by such collision. 
Our scheme, explained in details in the following, is based on a) allocating sequences such that such type of damaging collision occurs with low probability, and b) using a subspace estimation method that is robust to 
a small number of heavily interfered measurements (outliers). 

\subsection{Orthogonal latin squares-based SRS pilot hopping scheme} 

A latin square of order $N$ is an $N \times N$ array $\Am$ with elements in $[N] = \{1,\ldots, N\}$ such that every row and column contain all distinct elements. 
Two latin squares $\Am$ and $\Bm$ are said to be mutually orthogonal if the collection of elementwise pairs 
$\{(\Am(i,j), \Bm(i,j)) : i,j \in [N]\}$ contains all $N^2$ distinct pairs \cite{467960}.

\begin{example}  \label{latin-square}
	Consider $N = 5$ and the latin squares
	\begin{equation}
		\Am = \Resize{.24\linewidth}{\begin{bmatrix}
				\boxed{1} & 2 & 3 & 4 & 5 \\
				2 & 3 & 4 & 5 & \boxed{1} \\
				3 & 4 & 5 & \boxed{1} & 2  \\
				4 & 5 & \boxed{1} & 2 & 3 \\
				5 & \boxed{1} & 2 & 3 & 4
		\end{bmatrix}}  , \
		\Bm = \Resize{.24\linewidth}{\begin{bmatrix}
				\boxed{1} & 2 & 3 & 4 & 5  \\
				3 & 4 & 5 & 1 & \boxed{2} \\
				5 & 1 & 2 & \boxed{3} & 4  \\
				2 & 3 & \boxed{4} & 5 & 1  \\
				4 & \boxed{5} & 1 & 2 & 3 
		\end{bmatrix} }. \nonumber
	\end{equation}
	The positions marked by (say) 1 in $\Am$ (boxed positions) contains all elements $\{1,2,3,4,5\}$ in $\Bm$, and this holds for all symbols. Hence, $\Am$ and $\Bm$ are mutually orthogonal latin squares. \hfill $\lozenge$
\end{example}

The construction of families of $N - 1$ mutually orthogonal latin squares of order $N$ when $N$ is a prime power (i.e., $N = p^n$ for some prime $p$ and integer $n$) 
is well-known and given in \cite{467960}. In the proposed scheme, a latin square identifies $N$ hopping sequences as follows: we identify the rows of the 
latin square with the distinct subcarriers $f \in [N]$ used for SRS hopping (numbered from 1 to N without loss of generality) and 
the columns with the  time slots $s \in [N]$. 
Each latin square defines $N$ mutually orthogonal hopping sequences given by the sequence of positions $(f,s)$ corresponding to each integer $k \in [N]$. 
For example, in Example \ref{latin-square}, $N = 5$ users (say $k_1, k_2, \ldots, k_5$ are associated to the 5 mutually orthogonal hopping sequences
$\{ (f,s) : \Am(f,s) = n \}$, for $n = 1,2,\ldots, 5$ (e.g., user $k_1$ is associated to the sequence hopping over the boxed symbols in $\Am$ corresponding to index $n = 1$, and so on). 
If two users are associated to sequences in distinct orthogonal latin squares, they shall collide only in one hopping position. In other words, a user associated to latin square $\Am$ 
will collide on the $N$ hopping slots with each one of the $N$ users associated to the orthogonal latin square $\Bm$ (e.g., consider the boxed positions in $\Am$ and $\Bm$
in Example \ref{latin-square}, where user $k_1$ associated to $n = 1$ in $\Am$ collides with different users associated with hopping sequences $1, 5, 4, 3, 2$ 
in $\Bm$ in the order of consecutive slots $s = 1,2,3,4,5$ indicating the columns of the latin square array). 
This creates a good averaging of interference. In particular, we partition the network coverage area into 
hexagonal cells, and allocate the $N-1$ latin squares in a classical reuse of order $N-1$, such that adjacent cells have distinct orthogonal latin squares. 
The hexagonal cells have nothing to do with the user-centric clusters and are defined on a purely geometric basis. In practice, one may consider that the cell-free user-centric network
co-exists on a different frequency band with an existing LTE or 5GNR cellular network, and that the latin square allocation is done based on the cells of the cellular network.
Since cell-free user-centric networks are often regarded as non-stand-alone networks to provide high data rates to users already operating in a conventional network through (possibly non-adjacent) carrier aggregation, and since 5G allows the decoupling of the control and data planes \cite{3gpp23501}, such allocation can be easily accomplished through the cellular network.  
As a consequence of this geographic reuse of the family of orthogonal latin squares, users in the same hexagonal cell have no mutual SRS pilot collisions, and users 
in adjacent cells have at most one mutual SRS pilot collision.  
Notice that the good interference averaging property of orthogonal latin squares for slow hopping was already noticed in \cite{467960} 
in the context of 2G systems such as GSM. We conclude this section by saying that  in general the length of SRS hopping sequences $S$ may be larger than $N$. In this case, 
the time slot index $s$ is to be considered modulo $N$, that is, the sequence is periodically repeated. 
\begin{figure}[t]
	\centerline{
		\includegraphics[width=.99\linewidth]{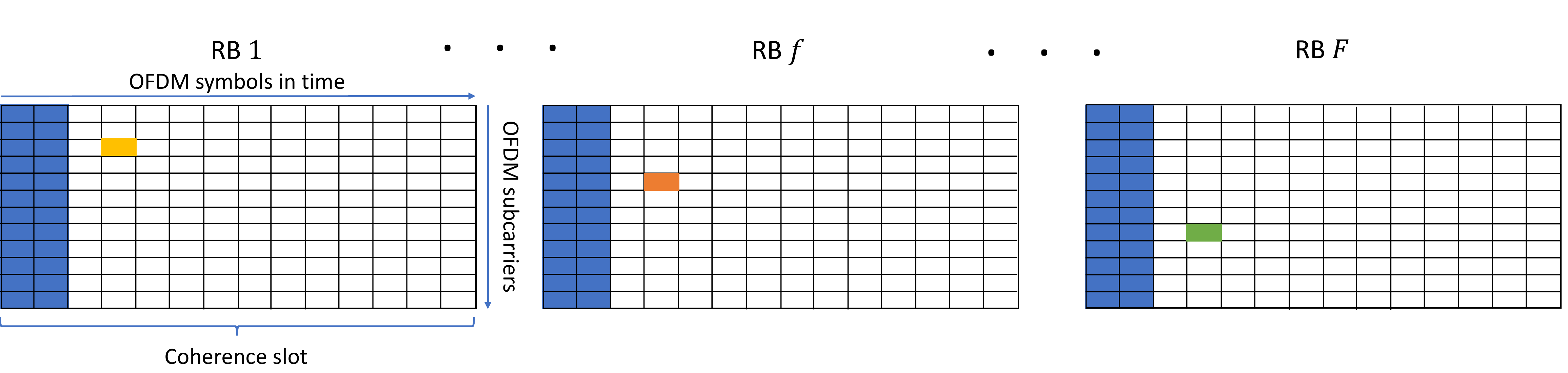} }
	\vspace{-.6cm}
	\caption{An example placement of DMRS (blue boxes, i.e., time-frequency symbols) and SRS pilots (orange, yellow, and green boxes), and data transmission (white boxes) on multiple RBs in one coherence slot. The yellow, orange, and green boxes are the SRS pilots of different UEs sent on different subcarriers.}
	\vspace{-.8cm}
	\label{fig:pilot_dimensions}
\end{figure}
\begin{remark}  {\bf The dimensionality of DMRS and SRS pilots.} 
	It is important to point out that the SRS pilots consume significantly fewer communication resources than the DMRS pilots. 
	While the SRS pilot is sent on one subcarrier per time slot and has dimension $1$, the DMRS pilots must be sent on  
	each coherence slot to allow the estimation of the UL instantaneous channel vectors, and requires $\taudmrs$ dimensions per slot, typically in the range $\taudmrs \in [20, 40]$. 
	The fraction of signal dimensions occupied by UE $k$'s DMRS pilots is 
	$\frac{\taudmrs} {T}$,  where $T$ denotes the signal dimensions of one coherence slot.  On the other hand, the SRS pilots require a set of $N$ frequency 
	domain symbols per slot, spanning the total system bandwidth. In general, the system bandwidth contains $F > 1$ (typically many tens) frequency-domain 
	RBs. It follows that the total cost of the SRS pilots is a fraction $\frac{N} {T F}$ signal dimensions per time slot. For example, assuming a channel bandwidth of 10 MHz and subcarrier spacing of 15 kHz in a system with Cyclic Prefix OFDM and 64QAM, we have $F = 52$ frequency-domain RBs according to a TDD UL reference measurement channel in \cite[Table A.2.3.8-1]{3gpp38101}. Let $N=61$ like in Section \ref{sec:simulations} such that $N/F = 61/52 \ll \taudmrs$, we can conclude that the 
	SRS pilots are negligible in terms of spectral efficiency cost.  Fig. \ref{fig:pilot_dimensions} shows an example for the proposed placement of DMRS and SRS pilots in a time-frequency grid, illustrating the negligible consumption of communication resources by the SRS pilots.
	\hfill $\lozenge$
\end{remark}

\subsection{Estimation via R-PCA}   \label{sec:RPCA}

Consider a generic RU $\ell$. The RU is aware of the SRS hopping sequence of all its associated users in $\Uc_\ell$. 
Focusing on some UE $k \in \Uc_\ell$, 
an RU $\ell \in \Cc_{k}$ collects all SRS pilot measurements corresponding to the hopping sequence of UE $k$. On slot $s \in [S]$ these are given by 
\begin{eqnarray}
	\yv_{\ell,k}^{\rm SRS}(s) 
	& = & \hv_{\ell,k} (s) + \sum_{i \neq k: t_i(s) = t_k(s) } \hv_{\ell,i} (s) + \widetilde{\zv}_{k,\ell}(s) \\
	& = & \hv_{\ell,k} (s) + \sum_{i \neq k: i \in \Ic^s_k(s) } \hv_{\ell,i} (s) +  \sum_{i \neq k: i \in \Ic^w_k(s) } \hv_{\ell,i} (s) + \widetilde{\zv}_{k,\ell} (s) \\
	& = & \hv_{\ell,k} (s) + \ev_{\ell,k} (s) +  \nv_{\ell,k} (s) ,
\end{eqnarray}
where $\widetilde{\zv}_{k,\ell}(s) \in \CC ^{M \times 1}$ with i.i.d. components $\Cc\Nc(0, \frac{1}{\SNR})$, the condition
$t_i(s) = t_k(s)$ indicates that the hopping sequence of user $i$ and of user $k$ collide at slot $s$, and 
the sets $\Ic^s_k(s)$ and $\Ic^w_k(s)$ contain the UEs colliding with UE $k$ with strong and weak LSFCs with respect to RU $\ell$, respectively. 
The term  $\ev_{\ell,k} (s) = \sum_{i \neq k: i \in \Ic^s_k(s) } \hv_{\ell,i} (s)$ accounts for the strong undesired signals (the so-called \textit{outliers}), while 
$\nv_{\ell,k} = \sum_{i \neq k: i \in \Ic^w_k(s) } \hv_{\ell,i} (s) + \widetilde{\zv}_{k,\ell} (s)$ includes noise and weak interference.
Stacking the $S$ SRS pilot observations as columns of an $M \times S$ array, we have
\begin{eqnarray}
	\Ym^{\rm SRS}_{\ell,k} & = & [\yv_{\ell,k}^{\rm SRS}(1) \ \yv_{\ell,k}^{\rm SRS}(2) \dots \yv_{\ell,k}^{\rm SRS}(S)] \label{Y_concat} 
	= \Hm_{\ell,k} + \Nm_{\ell,k} + \Em_{\ell,k}, 
\end{eqnarray}
where $\Hm_{\ell,k}$, $\Nm_{\ell,k}$ and $\Em_{\ell,k}$ are given by the analogous stacking of vectors $\hv_{\ell,k} (s), \ev_{\ell,k} (s)$ and $ \nv_{\ell,k} (s)$, respectively. 
Because of the orthogonal latin squares hopping patterns, it is expected that 
$\Em_{\ell,k}$ is column-sparse to a certain degree.  This is equivalent to the noise plus outliers model in \cite{6126034}. 
The R-PCA algorithm in  \cite{6126034} aims at detecting outliers, i.e., the non-zero columns of $\Em_{\ell,k}$, and at estimating the subspace of $\Hm_{\ell,k}$, which eventually is the desired channel subspace
of UE $k$ at RU $\ell$. Fixing some 
$\epsilon > 0$ and $\lambda > 0$, 
the algorithm solves the  convex problem
\begin{eqnarray}
	\underset{ \Hm_{\ell,k}, \Em_{\ell,k} }{\text{minimize}} \;\;\;  \lVert \Hm_{\ell,k} \rVert_* + \lambda \lVert \Em_{\ell,k} \rVert_{2,1}, & &  
	\text{subject to:} \;\;\; \lVert \Ym^{\rm SRS}_{\ell,k} - \Hm_{\ell,k} - \Em_{\ell,k} \rVert_F \leq \epsilon,  \label{pca_opt_problem} 
\end{eqnarray}
where $\lVert \cdot \rVert_*$, $\lVert \cdot \rVert_F$, and $\lVert \cdot \rVert_{2,1}$ denote the nuclear norm, the Frobenius norm, and the sum of the $\ell_2$ column norms of a matrix, respectively. The Lagrangian function of (\ref{pca_opt_problem}) is given by 
\begin{align}
	\Lc \left (\Hm_{\ell,k}, \Em_{\ell,k}, \lambda, \mu \right ) = \lVert \Hm_{\ell,k} \rVert_* + \lambda \lVert \Em_{\ell,k} \rVert_{2,1} + \mu \left ( \lVert \Ym^{\rm SRS}_{\ell,k} - \Hm_{\ell,k} - \Em_{\ell,k} \rVert_F - \epsilon \right ),
\end{align}
where $\mu$ is a Lagrange multiplier.
Therefore, the corresponding unconstrained convex minimization problem is given as
\begin{align}
	\underset{ \Hm_{\ell,k}, \Em_{\ell,k} }{\min}  \; \underset{\mu \geq 0}{\max} \hspace{.2cm} \Lc \left (\Hm_{\ell,k}, \Em_{\ell,k}, \lambda, \mu \right ). \label{pca_convex}
\end{align}
We employ the algorithm proposed in \cite{6126034} to approach (\ref{pca_convex}), which returns estimates $\widehat{\Hm}_{\ell,k}$ and $\widehat{\Em}_{\ell,k}$ 
of the channel and outliers matrix, respectively. 
The parameter $\lambda$ is given as an input to the algorithm and is optimized empirically. 
The algorithm does not require $\epsilon$ to be specified and treats the observation matrix as in the noiseless case. Note that $\epsilon$ however is important for the analytical results of the algorithm. The Lagrange multiplier $\mu$ is optimized by the algorithm via primal-dual iterations. We refer to \cite{6126034} for more details of the employed R-PCA algorithm.

From the SVD $\widehat{\Hm}_{\ell,k} = \widehat{\Um}\widehat{\Sm} \widehat{\Vm}^\text{H}$, we estimate the subspace by considering the left singular vectors (columns of $\widehat{\Um}$) corresponding to the dominant singular values. One approach to find the number of dominant singular values is to find the index at which there is the largest difference (gap) between consecutive singular values.
Let $\widehat{\Fm}_{\ell,k}^{\rm PCA} = \widehat{\Um}_{:,1:r^{\rm PCA}}$ denote the tall unitary matrix obtained by selecting the dominant $r^{\rm PCA}$ left eigenvectors as explained above.  We can further post-process the subspace estimate by imposing that its basis vectors are DFT columns. 
As anticipated before, this is motivated by the fact that, for large Toeplitz matrices, the eigenvectors are closely approximated by DFT vectors \cite{adhikary2013joint}. 
Let $\Fm$ denote the $M \times M$ unitary DFT matrix with $(m,n)$-elements
	$[\Fm]_{m,n} = \frac{e^{-j\frac{2\pi}{M} mn}}{\sqrt{M}} , \; m, n  = 0,1,\ldots, M-1$.
The identification of the best fitting DFT columns to the R-PCA estimated subspace can be done greedily by finding one by one the $r^{\rm PCA}$ columns $\Fm_{:,i}$ of $\Fm$ 
that maximize the quantity
\begin{eqnarray}
	\Fm_{:,i}^\herm \widehat{\Fm}_{\ell,k}^{\rm PCA} (\widehat{\Fm}_{\ell,k}^{\rm PCA})^\herm \Fm_{:,i}.
\end{eqnarray}
We denote the selected set of column indices as $\widehat{\Sc}_{\ell,k}^{\rm PP}$, such that the corresponding estimated projected PCA (PP) 
subspace is given by $\widehat{\Fm}_{\ell,k}^{\rm PP} = \Fm(: , \widehat{\Sc}_{\ell,k}^{\rm PP})$ (the columns of $\Fm$ indexed by $\widehat{\Sc}_{\ell,k}^{\rm PP}$). The estimated channel covariance matrix for a given subspace estimate 
$\widehat{\Fm}_{\ell,k}$ is given by 
\begin{eqnarray}
	\Sigma_{\hv}(\widehat{\Fm}_{\ell,k}) = \frac{\beta_{\ell,k} M}{r^{\rm PCA}} \widehat{\Fm}_{\ell,k} \widehat{\Fm}_{\ell,k}^\herm, 
\end{eqnarray} 
where $\widehat{\Fm}_{\ell,k} = \widehat{\Fm}_{\ell,k}^{\rm PCA}$ or $\widehat{\Fm}_{\ell,k} = \widehat{\Fm}_{\ell,k}^{\rm PP}$, depending on the considered case. 

In order to evaluate the quality of a generic subspace estimate $\widehat{\Fm}_{\ell,k}$, we consider the power efficiency (PE) defined by 
$E_{\rm PE}(\widehat{\Fm}_{\ell,k}) = \frac{\trace\left(\Sigma_{\ell,k} \Sigma_{\hv}(\widehat{\Fm}_{\ell,k}) \right)}{\trace\left(\Sigma_{\ell,k} \Sigma_{\ell,k}\right)}  \in [0,1] .   \label{PE}$
The PE
measures how much power from the desired channel in (\ref{chest1}) is captured in the channel estimate. 
Additionally, we consider the normalized Frobenius-norm error given by
$
	E_{\rm NF}(\widehat{\Fm}_{\ell,k}) = \frac{ \lVert \Sigma_{\ell,k} - \Sigma_{\hv}(\widehat{\Fm}_{\ell,k}) \rVert_\text{F} }{\lVert \Sigma_{\ell, k} \rVert_\text{F}}. \label{NF}
	$

\section{Simulations}  \label{sec:simulations}

We consider a square with  area of $A = 2 \times 2\;\; {\rm km}^2$ with a torus topology to avoid boundary effects. 
The LSFCs (including distance-dependent pathloss, blocking effects, and shadowing) are given by the 3GPP urban microcell street canyon pathloss model from \cite[Table 7.4.1-1]{3gpp38901}, which differentiates between UEs in line-of-sight (LOS) and non-LOS (NLOS). The probability of LOS is distance-dependent and given in \cite[Table 7.4.2-1]{3gpp38901}. 
	A log-normal Gaussian random variable with different parameters for LOS and NLOS is added to the deterministic pathloss model to account for shadow fading.
	For the spatial correlation between the channel antenna coefficients, we consider a simple directional channel model defined as follows. 
	Consider the angular support $\Theta_{\ell,k} = [\theta_{\ell,k} - \Delta/2, \theta_{\ell,k} + \Delta/2]$ centered at angle $\theta_{\ell,k}$ of the LOS 
	between RU $\ell$ and UE $k$ (with respect to the RU boresight direction), with angular spread $\Delta$. 
	Then, we let
	\begin{equation} 
		\hv_{\ell,k} = \sqrt{\frac{\beta_{\ell,k} M}{|\Sc_{\ell,k}|}}  \Fm_{\ell,k} \nuv_{\ell, k}, \label{channel_model}
	\end{equation}
	where the index set  $\Sc_{\ell,k} \subseteq \{0,\ldots, M-1\}$ includes all integers $m$ such that $2\pi m/M \in \Theta_{\ell,k}$
	(where angles are taken modulo $2\pi$), where
	$\Fm_{\ell,k}$ is the submatrix extracted from $\Fm$ by taking the columns indexed by $\Sc_{\ell,k}$,  
	and  $\nuv_{\ell,k}$ is an $|\Sc_{\ell,k}| \times 1$ i.i.d. Gaussian vector with components 
	$\sim \Cc\Nc(0,1)$. It follows that  $\hv_{\ell,k}$ is a Gaussian zero-mean random vector confined in the subspace spanned by the columns 
	of $\Fm_{\ell,k}$.  Notice that this channel model corresponds to the single ring of scatterers located around the UE \cite{adhikary2013joint}, with suitable 
	quantization of the angle domain according to the $M$-array resolution limit. While the one-ring scattering model may be restrictive and it is used in the simulations for simplicity, the fact that the dominant channel subspace is spanned by a collection of (possibly non-adjacent) columns of an $M\times M$ unitary DFT matrix is actually general enough for large ULAs and UPAs. The angular support $\Sc_{\ell,k}$ imposes a simplified but meaningful form of geometric consistency in the antenna correlation: if two users are located in the same $\Delta$-wide angle with respect to an RU, their channel vectors will have identical subspace and therefore identical covariance matrix. 
	In particular, if the spanned subspace has dimension 1 (i.e., generated by a 
	single DFT column as in a LOS situation), the two users will have co-linear channels with respect to the RU. 
	We believe that this spatial geometry consistency  effect is much more realistic than the traditional i.i.d. channel models used in 
	early (cell-free) massive MIMO works.

We consider a system with $K = 100$ UEs and $LM = 640$ total antennas at all RUs, where the level of antenna distribution, i.e., the values of $L$ and $M$, varies. 
UEs and RUs are randomly and independently distributed in the area $A$.  We use $\Delta = \pi/8$, the maximum cluster size of $Q = 10$ RUs serving one UE, and the SNR threshold in (\ref{eq:snr_threshold}) is set to $\eta = 1$.
A bandwidth of $W = 10\text{ MHz}$ and noise with power spectral density of $N_0 = -174 \text{ dBm/Hz}$ is considered. 
The UL energy per symbol is chosen such that $\bar{\beta} M \SNR = 1$ (i.e., 0 dB), when the expected pathloss $\bar{\beta}$ with respect to LOS and NLOS is calculated for distance 
	$3 d_L$, where $d_L = \sqrt{\frac{A}{\pi L}}$ is the radius of a disk of area equal to $A/L$. In this way, the UL Tx power of all UEs is dependent on the RU density 
and number of RU antennas to achieve a certain level of overlap of the RUs' coverage areas, such that each UE is likely to be associated to several RUs.
Table \ref{table_noise_snr} summarizes the specific values following this approach for different system configurations. The actual (physical) UE transmit power
is obtained as $P_{\rm tx}^{\rm ue} = P^{\rm ue}W$ and it is expressed in dBm.   Table \ref{table_noise_snr} shows that, for our system parameters, the resulting
UE transmit power is in the range of approximately $13 \text{ to }19 \text{ dBm}$, which is realistic. 
\begin{table}[t]
	\caption{Parameters for different system configurations.}
	\vspace{-.6cm}
	\begin{center}
		\begin{tabular}{ | l | c | c | c |}
			\hline
			{\{$L, M$\}} & $d_L\text{[m]}$ & $\bar{\beta}\text{[dB]}$ & {$P^{\rm ue}_{\rm tx} \text{[dBm]}$} \\  \hline
			\{10, 64\} & 356.825 & -140.849 & 18.7872 \\ \hline 
			\{20, 32\} & 252.313 & -135.334 & 16.282 \\ \hline 
			\{40, 16\} & 178.412 & -129.728 & 13.687 \\ \hline 
		\end{tabular}
		\label{table_noise_snr}
	\end{center}
	\vspace{-1cm}
\end{table}
\subsection{Evaluation of the R-PCA algorithm for system simulations}

As the computational complexity of the R-PCA is relatively high, we first show how the R-PCA algorithm can be replaced in the system simulations with a distribution-based method. This approach generates subspace estimates from the empirical distributions for multiple channel gain ranges that describe the probability of the possible outputs of the R-PCA algorithm. 
We validate our approach by comparing the distribution-based method with the actual R-PCA in terms of the subspace estimation performance measures defined in Section \ref{sec:RPCA}. We only consider the outputs of the PP estimates for the distribution-based approach, since in any case this yields the best results 
and it is motivated by the already mentioned fact that, for large $M$, the channel covariance eigenvectors are well approximated by columns of the DFT matrix. 

By construction,  the subspace $\Fm_{\ell,k}$ is uniquely identified by the set of adjacent integers $\Sc_{\ell,k} \subseteq \{0, \ldots, M-1\}$ taken modulo $M$. 
Let $\widehat{\Sc}_{\ell,k}^{\rm PP}$ denote the estimation of the true subspace index $\Sc_{\ell,k}$ obtained from the R-PCA scheme followed by projection on the DFT columns as described in Section \ref{sec:RPCA}. By the circular symmetry of the one-ring scattering model, it follows that the posterior distribution of 
$\widehat{\Sc}_{\ell,k}^{\rm PP}$ given $\Sc_{\ell,k}$ is invariant with respect to shifts modulo $M$, in other words, 
letting $P(\widehat{\Sc}_{\ell,k}^{\rm PP} | \Sc_{\ell,k})$ such posterior distribution, for any integer $m$ we have
\[   P(\widehat{\Sc}_{\ell,k}^{\rm PP} + m | \Sc_{\ell,k} + m) = P(\widehat{\Sc}_{\ell,k}^{\rm PP} | \Sc_{\ell,k}). \]
where the addition operation in the subspace index set is modulo $M$ (cyclic shift). 
In addition, we noticed that given a certain system random geometry (i.e., the coverage area $A$, number of RUs $L$ and UEs $K$, 
with random placement thereof, number of per-RU antennas $M$, pathloss law, and angular spread $\Delta$), the most important parameter
that determines the statistics of the subspace estimate for UE $k$ at RU $\ell$  is the LSFC $\beta_{\ell,k}$. 
Hence, our approach consists of  generating by Monte Carlo simulation a large set of pairs $(\Sc_{\ell,k} , \widehat{\Sc}_{\ell,k}^{\rm PP})$, group them in bins 
according to suitably discretized grid values  for  $\beta_{\ell,k}$, and use the rotation symmetry said above, in order to obtain 
the empirical conditional distribution of the estimated subspace $\widehat{\Sc}_{\ell,k}^{\rm PP}$ given $\Sc^{\rm ref}_{\ell,k} = \{0,1,\ldots, \lfloor \frac{\Delta M}{2\pi} \rfloor -1 \}$ and 
$\beta_{\ell,k} \in \delta\beta_i$, where $\delta\beta_i$ denotes the $i$-th interval of the discretized grid. 

Denoting by  $P(\widehat{\Sc}_{\ell,k}^{\rm PP} | \Sc^{\rm ref}_{\ell,k}, \beta_{\ell,k} \in \delta\beta_i)$ the generated family of empirical conditional distributions (for all intervals of the pathloss), we run our large-scale system simulations by replacing the actual R-PCA PP approach with the following generative rule: 
for each $\ell \in [L]$ and $k \in \Uc_\ell$, let $m_{\ell,k}$ the left-most subspace index in $\Sc_{\ell,k}$ and let $\beta_{\ell,k}$ the LSFC. 
Suppose that $\beta_{\ell,k} \in \delta\beta_i$,  then  generate the estimated subspace $\widehat{\Sc}_{\ell,k}^{\rm PP}$ by sampling from the 
corresponding $i$-th empirical distribution and cyclic-shifting  the obtained set of indices by $m_{\ell,k}$ (i.e., adding $m_{\ell,k}$ modulo $M$).

\begin{figure}[t!]
	\centerline{\includegraphics[width=7.8cm]{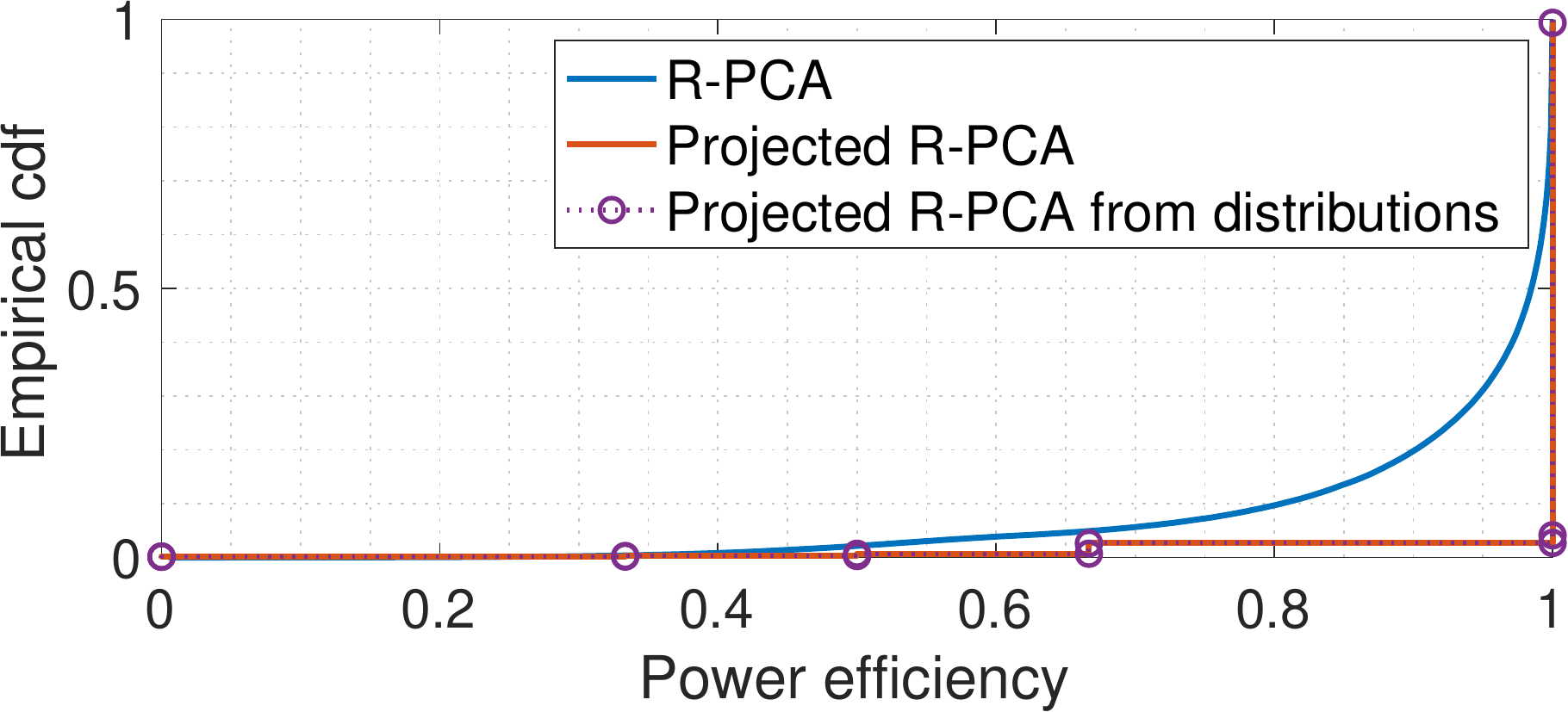} \hspace{.6cm} \includegraphics[width=7.8cm]{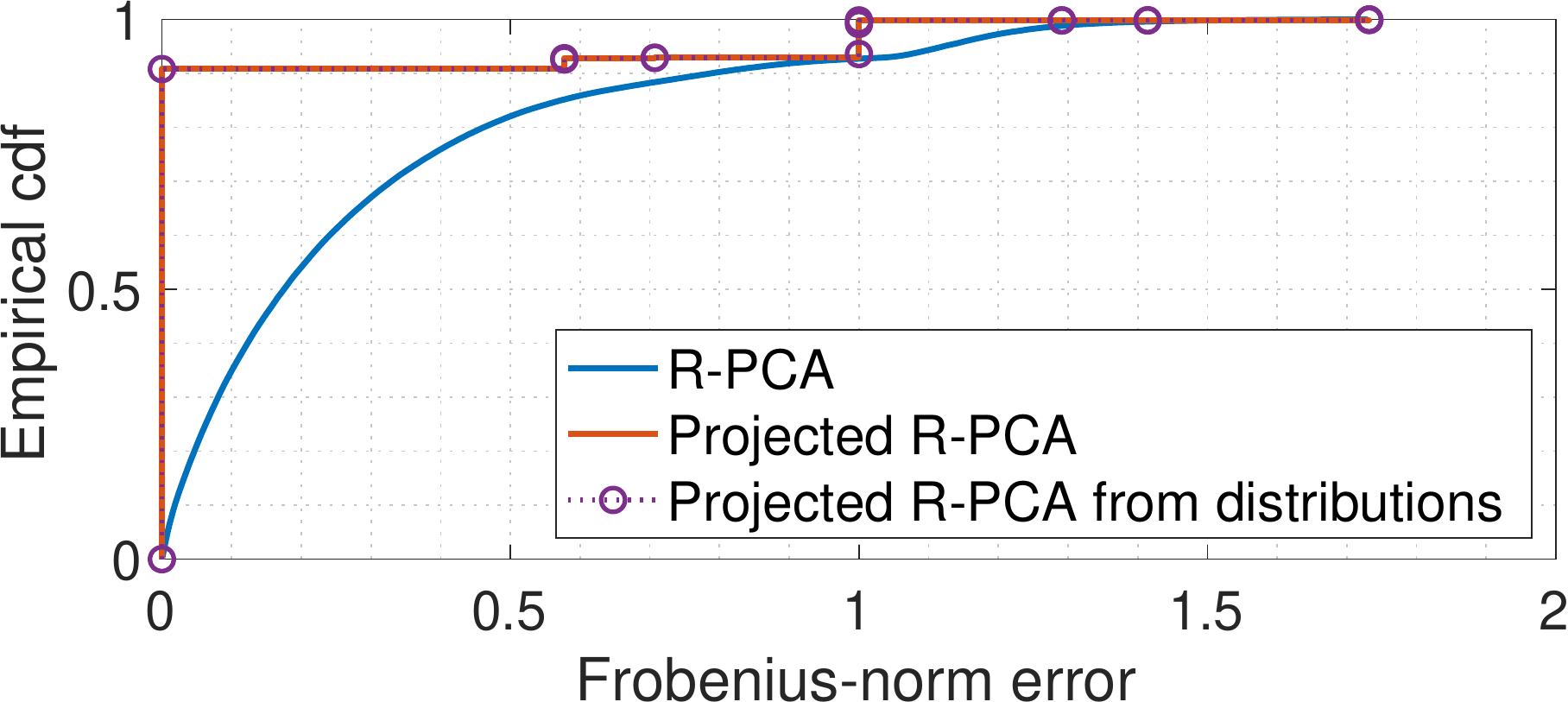}}
	\vspace{-.6cm}
	\caption{The power efficiency and Frobenius-norm error of the R-PCA  subspace estimates and of the corresponding distribution-based generated estimates.}
	\label{subspace_estimation_results_pca}
	\vspace{.3cm}
	\centerline{\includegraphics[width=7.8cm]{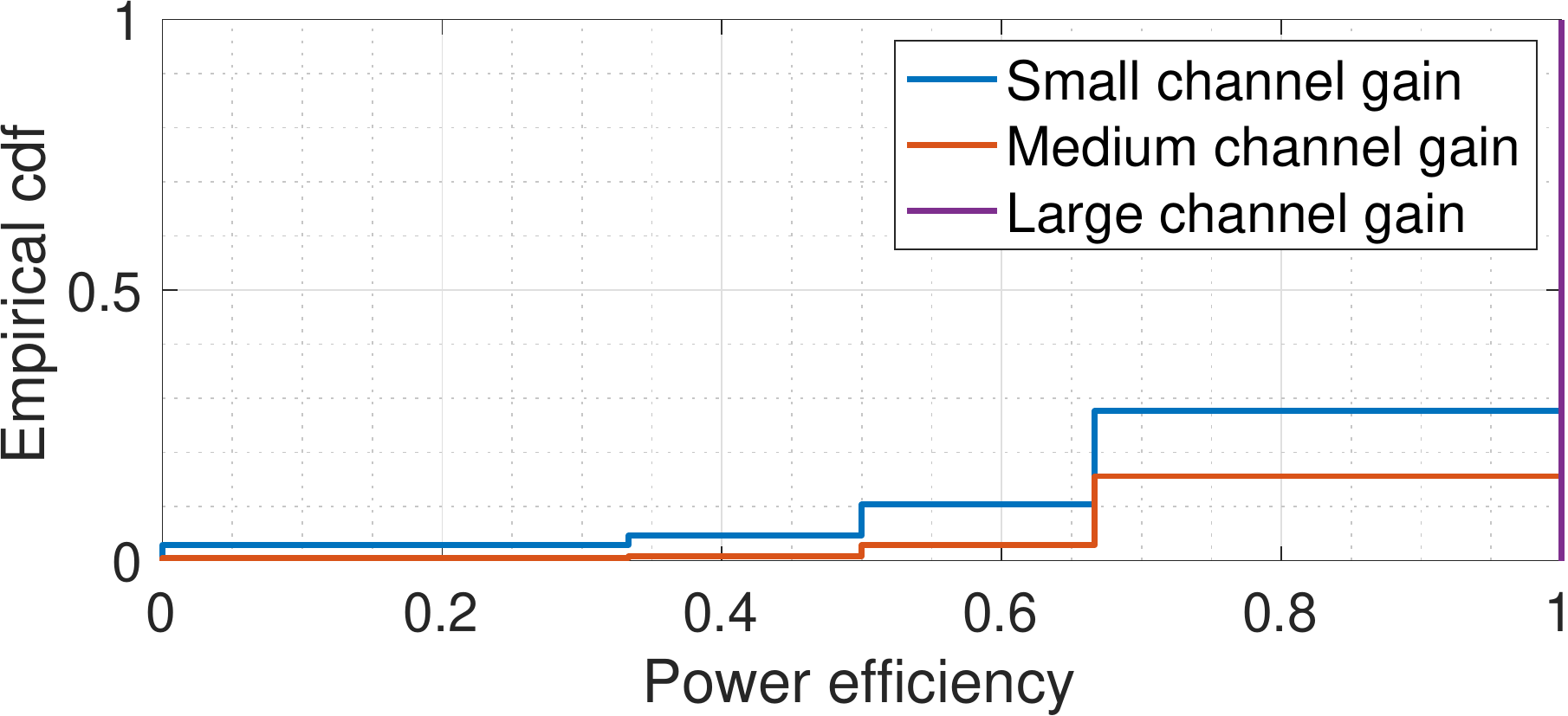} \hspace{.6cm} \includegraphics[width=7.8cm]{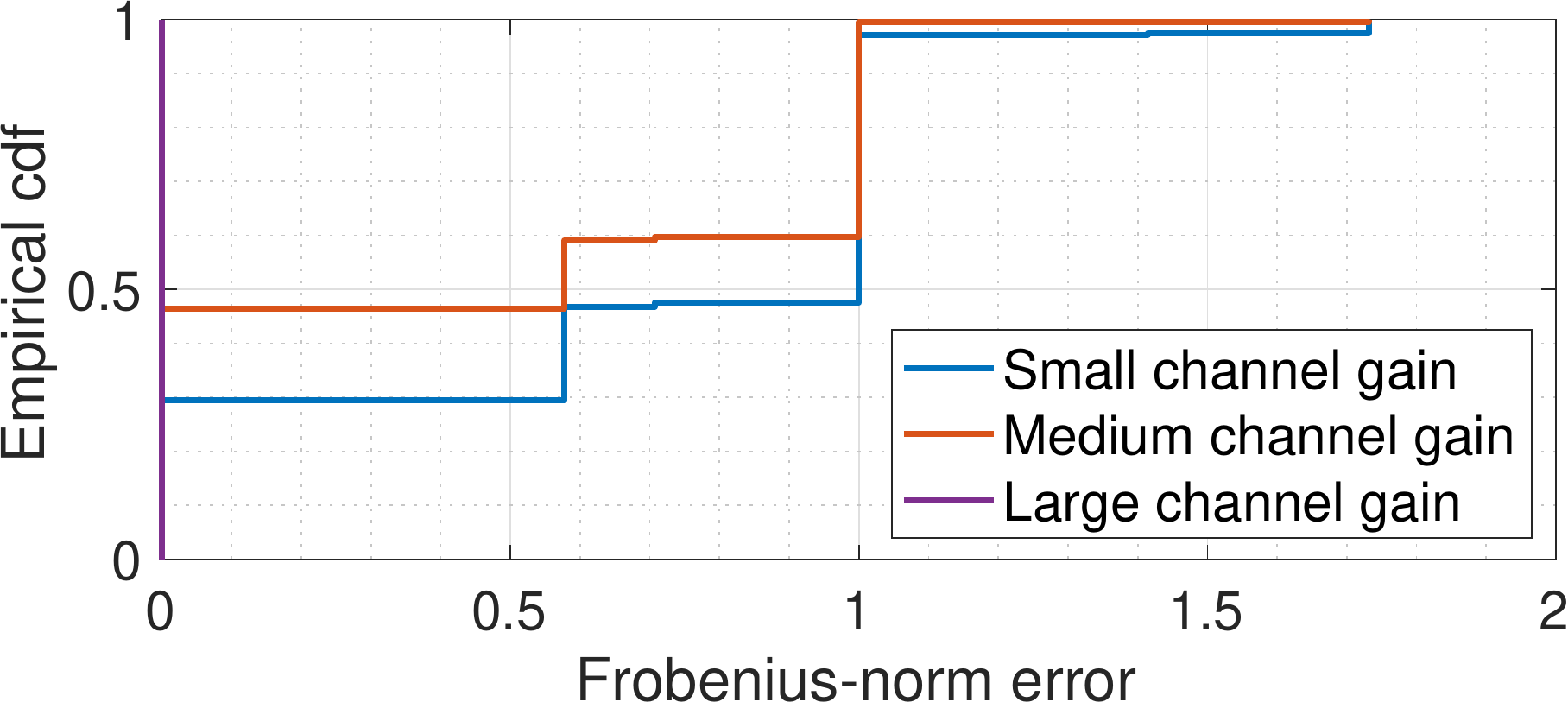}}
	\vspace{-.6cm}
	\caption{The power efficiency and Frobenius-norm error of the projected estimates with the R-PCA algorithm in different channel gain ranges, where $L=40$, $M=16$ and $N = 61$. ``Small channel gain'' accounts for the dataset of the channel gains in the 5th percentile, ``Medium channel gain'' for the channel gains between the 5th and 10th percentile, ``Large channel gain'' for the channel gains above the 90th percentile.}
	\vspace{-.8 cm}
	\label{compare_rpca_pdf_vs_beta_ranges}
\end{figure}

We evaluate the subspace estimation accuracy of the R-PCA algorithm and the distribution-based approach in a system with $L=40$ RUs, $M=16$ RU antennas and $K=100$ UEs. 
We let the SRS pilot sequences consist of $S=200$ samples and hop over $N=61$ distinct subcarriers. Note that we have observed no significant improvements for larger $N$, while for smaller $N$ the estimation accuracy is degraded. Therefore, for the sake of brevity we have kept $N = 61$ fixed throughout our results.
Fig. \ref{subspace_estimation_results_pca} shows the PE of the subspace estimates of the R-PCA algorithm and the distribution-based approach for 300 independent topologies (random uniform placement of RUs and UEs), where the PE of both the output of the R-PCA algorithm and of the projected estimate are given.
As the R-PCA estimates the subspace without the knowledge that it is constructed by a set of columns of the DFT matrix, the results of the R-PCA outputs follow a continuous distribution. 
The results of the post-processed estimates follow approximately the distribution of the R-PCA-based estimates, with the difference that due to the post-processing the set of possible outcomes is discrete. 
The approach using empirical conditional distributions given the channel gain of an RU-UE pair yields very similar results compared to the post-processed estimates of the R-PCA algorithm.

This legitimates the use of the empirical conditional distributions in the system simulations, and the proposed approach shows a method to run system simulations containing possibly hundreds of RUs and UEs with highly reduced computational complexity compared to employing the R-PCA algorithm for each RU-UE pair.

\begin{remark}  {\bf Construction of the datasets for the distribution-based subspace estimation simulation.} 
	For each system configuration of $L, M, K, N$, we simulated 300 independently generated network topologies with random placement of RUs and UEs. For each topology we run the R-PCA algorithm with DFT projection for all UEs $k$ and all the associated RUs $\ell \in \Cc_k$, according to the SRS pilot allocation and cluster formation schemes described in this paper. 
	For each such $(\ell,k)$ pair we collected the true and estimated channel subspace index sets $\Sc_{\ell,k}$ and $\widehat{\Sc}_{\ell,k}^{\rm PP}$ , respectively, and
	the value of the LSFC $\beta_{\ell,k}$. Then, we partitioned the obtained data set according to  a grid of 20 intervals of the LSFC values, where the intervals are defined by successive 5th percentiles.  For example, the first subset contains all pairs  $(\Sc_{\ell,k}, \widehat{\Sc}_{\ell,k}^{\rm PP})$ for which the corresponding $\beta_{\ell,k}$
	is in  the lower 5\% of the values of $\beta$. In this way, each subset contains the same number of elements. Finally, we used the subsets to construct the 
	family of empirical conditional distributions $P(\widehat{\Sc}_{\ell,k}^{\rm PP} | \Sc^{\rm ref}_{\ell,k}, \beta_{\ell,k} \in \delta\beta_i)$, for $i = 1, \ldots, 20$. 
	As said before, this is motivated by the fact that the subspace estimation accuracy of the R-PCA algorithm is highly dependent on the value of the 
	LSFC between the 
	UE-RU pair, as Fig. \ref{compare_rpca_pdf_vs_beta_ranges} illustrates. 
	\hfill $\lozenge$
\end{remark}

\subsection{System performance with subspace estimates}

For the evaluation of the data rate and spectral efficiency (SE), we generate 100 topologies for each set of parameters, 
and compute the expectation in (\ref{ergodic_rate_ul}) and (\ref{ergodic_rate_dl}) by Monte Carlo averaging with respect to the channel vectors. 
The UL (same for DL) SE for UE $k$ is given by 
\begin{equation}
	{\rm SE}^{\rm ul}_k =  (1 - \taudmrs/T) R_k^{\rm ul}.
\end{equation}
We evaluate the UL and DL system performance achieved by the 
	combining and precoding schemes described in Section \ref{sec:schemes}, with DL power allocation from UL-DL nominal SINR duality.

\begin{remark}\ {\bf Comparison to a state-of-the-art local precoding method.}
	The cluster-level precoding schemes are compared to a state-of-the-art local precoding scheme, local zero-forcing (LZF), employed as follows. For the case that $|\Uc_\ell| > M$ as the outcome of the clustering process, the RU $\ell$ selects at most $M$ of its $|\Uc_\ell| $ associated UEs with the largest LSFCs and linearly independent channel vectors. The selected UEs are served by RU $\ell$, while the other UEs are dropped.
	We use proportional power allocation (PPA) with regard to the LSFCs such that 
	\begin{eqnarray}
		q_{\ell, k} = P^{\rm RU} \frac{\beta_{\ell,k}}{ \sum_{j \in \Uc_\ell} \beta_{\ell, j} }, \ \forall k \in \Uc_\ell , \label{eq:ppa}
	\end{eqnarray}
	for all $(\ell,k) \in \Ec$, where $q_{\ell, k}$ and $P^{\rm RU}$ denote the transmit power allocated at RU $\ell$ to UE $k$ and the DL power budget at each RU, respectively. For $k \notin \Uc_\ell$ (including the dropped UEs by the LZF UE selection), we have $q_{\ell, k} = 0$. The SINR with distributed DL power allocation becomes
	\begin{eqnarray}
		\SINR^{\rm dl-dist}_k = \frac{ \sum_{\ell \in \Cc_k} |\hv_{\ell,k}^\herm \uv_{\ell,k}|^2 q_{\ell,k}}{\SNR^{-1} + \sum_{j\neq k}  \sum_{\ell \in \Cc_j} |\hv_{\ell,k}^\herm \uv_{\ell,j}|^2 q_{\ell,j} }. \label{DL-dist} 
	\end{eqnarray}
	For a fair comparison with the cooperative power allocation schemes, we define $P^{\rm RU} = K/L$, such that the sum DL Tx power of all RUs is the same for all schemes. We also compare with local partial zero-forcing (LPZF) from \cite{9069486} and PPA, where each RU divides its set of associated UEs in two subsets, one served with ZF, one with maximum ratio transmission. We refer to \cite{9069486} for details of the LPZF scheme.
	\hfill $\lozenge$
\end{remark}

\begin{figure}[t]
	\centerline{\includegraphics[width=7.8cm, trim=.72cm 0 1.59cm .96cm,clip]{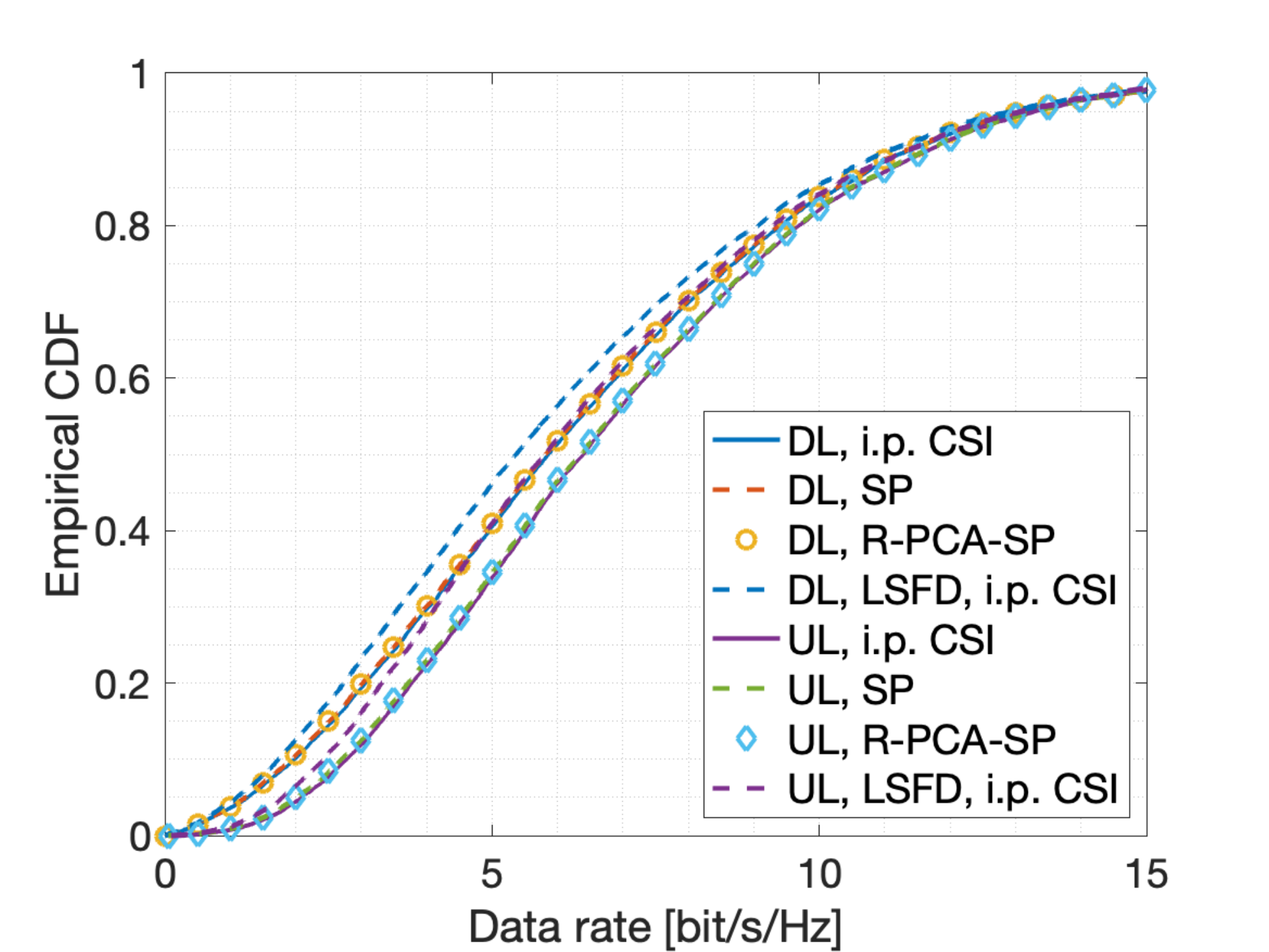} \hspace{.0cm} 
		\includegraphics[width=7.8cm]{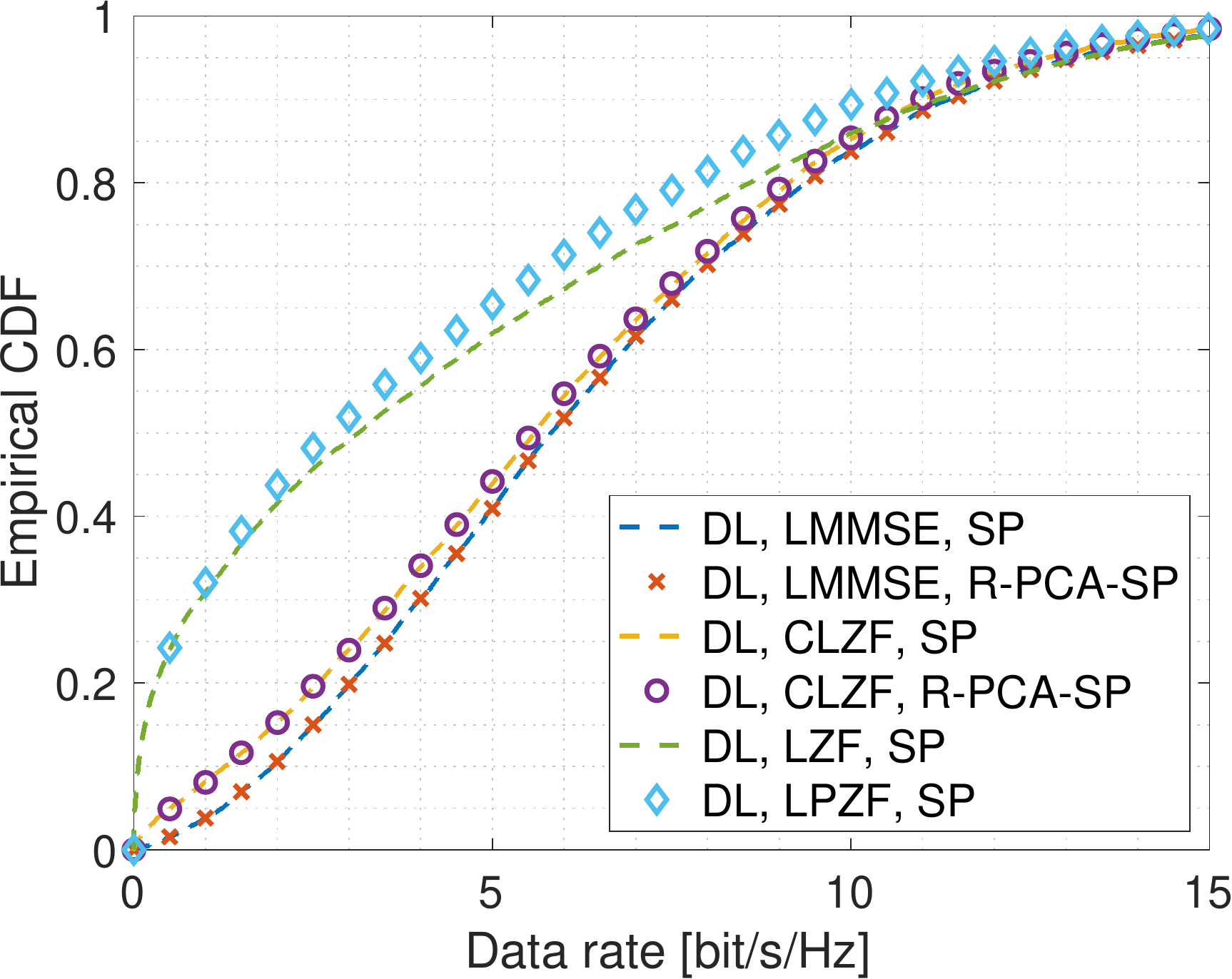} }
	\vspace{-.6 cm}
	\caption{The UL and DL data rates for LMMSE combining with DL power allocation from UL-DL duality (left). The DL data rates for different precoding schemes, where LMMSE/CLZF do power allocation from duality and LZF/LPZF use PPA (right).}
	\vspace{-.8 cm}
	\label{cdfplots_UL_DL_L40_pca_vs_sp_ideal}
\end{figure}

We consider a system with $L=40$, $M=16$, $K=100$, $\taudmrs = 20$, and RBs of dimension $T = 200$ symbols. 
Fig. \ref{cdfplots_UL_DL_L40_pca_vs_sp_ideal} shows that data rates with SP channel estimation based on R-PCA subspace estimates (denoted by ``R-PCA-SP'') can closely approximate the UL data rates assuming perfect subspace knowledge (indicated by ``SP'') and ideal partial channel knowledge (``i.p. CSI''). 
These results demonstrate that the problem of pilot contamination can essentially be solved in cell-free massive MIMO systems if the UL SRS pilot sequences are appropriately designed. Even the bad subspace estimates do not degrade the data rates significantly, since these occur for RU-UE pairs with a relatively low channel gain, whose contribution to the UE's data rate is small.

The left plot of Fig. \ref{cdfplots_UL_DL_L40_pca_vs_sp_ideal} shows the data rates achieved with LMMSE combining in the UL, and with a reuse of the combining vectors and power allocation from duality in the DL. The proposed UL-DL nominal SINR duality yields virtually symmetric UL and DL data rates, respectively. For the sake of comparison, we include the results achieved by LSFD (see {\em Remark \ref{remark_lsfd}}) with ideal partial CSI for both the UL and DL,
	and we observe a slight degradation compared to the proposed scheme, due to the usage of long term statistics instead of instantaneous channel information.
	In the right plot, we observe that the cooperative DL schemes outperform LZF and LPZF, where LMMSE precoding provides slightly higher performance at UEs with low data rates compared to CLZF. The superiority of the cooperative schemes is explained by the enhanced system information level available at the RU clusters compared to single RUs. In addition, due to the PPA employed by LZF and LPZF, the RUs transmit with little power to UEs with a small channel gain, leading to UEs with very low data rates compared to the cooperative schemes.
For a more thorough comparison of LZF and LPZF, we refer to \cite{kddi_uldl_precoding}.


\subsection{Uplink system-level performance for different system configurations}

\begin{figure}[t]
	\centerline{\includegraphics[width=7.8cm]{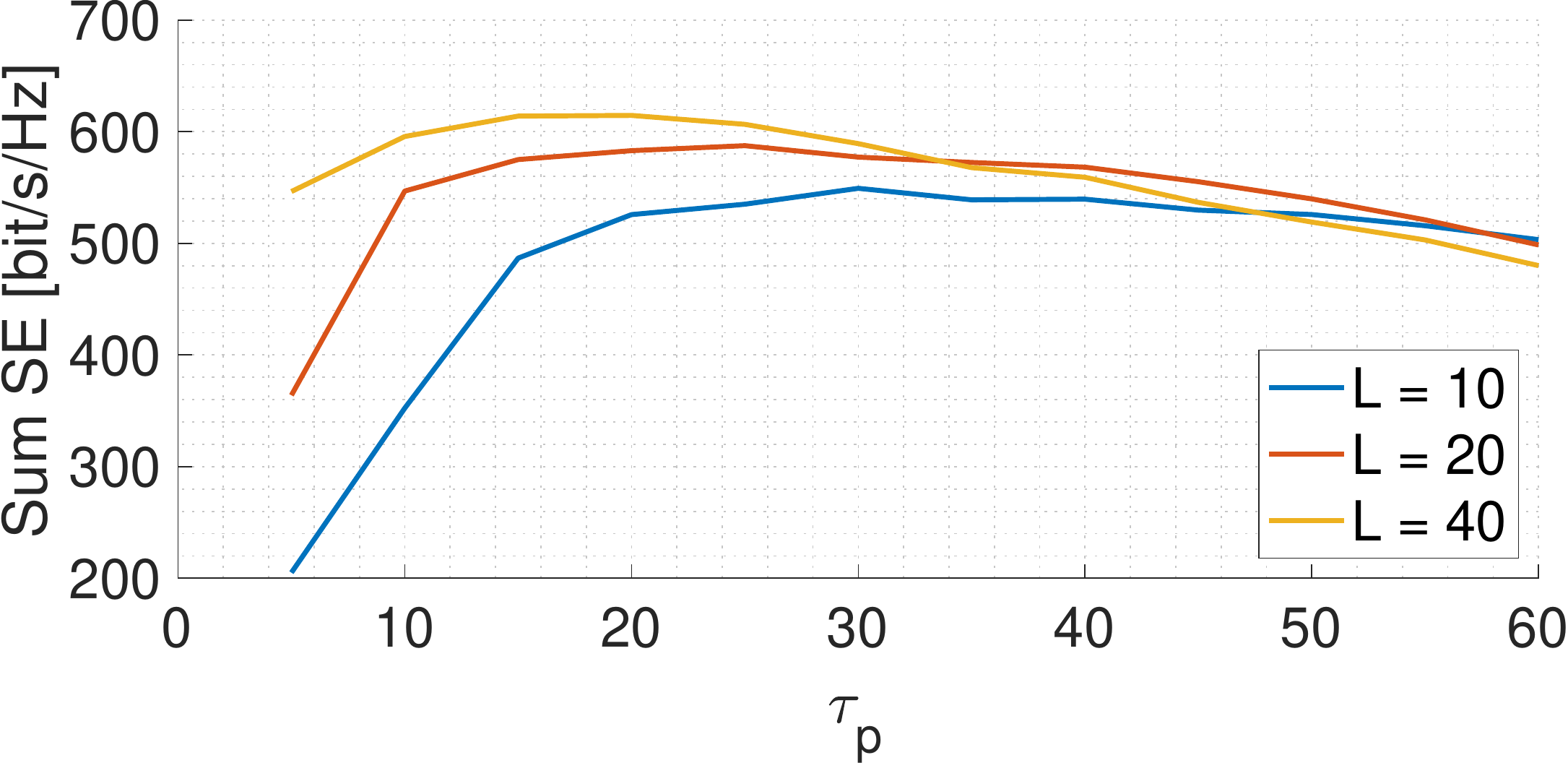} \hspace{.0cm} 
		\includegraphics[width=7.8cm]{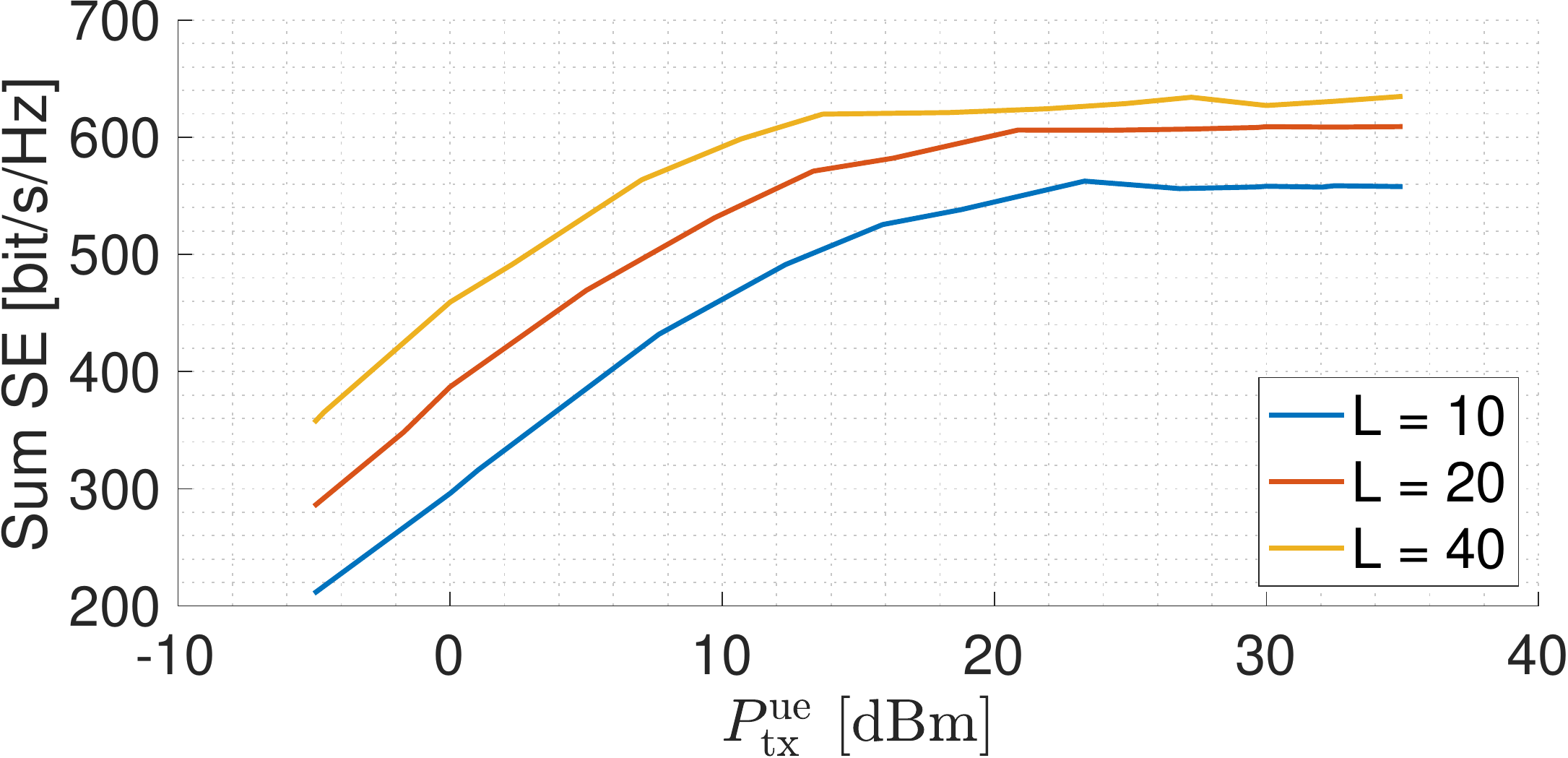} }
	\vspace{-.6 cm}
	\caption{The UL sum SE with LMMSE combining vs. $\taudmrs$ for different system configurations (left). The UL sum SE with LMMSE combining vs. $P^{\rm ue}_{\rm tx}$ for different system configurations (right).}
	\vspace{-.8 cm}
	\label{sum_se_vs_taup_and_tx}
\end{figure}

We compare the UL sum SE for different antenna distributions using LMMSE combining and subspace projection channel estimation under the assumption of perfect subspace knowledge (as we know that estimated subspaces closely approach the performance of perfect subspace knowledge), where $K=100$, $L=\{10,20,40\}$ and $M=\{64,32,16\}$. Fig. \ref{sum_se_vs_taup_and_tx} shows that the most distributed configuration with $L=40$ achieves the highest UL sum SE with optimized pilot dimensions, while $L=10$ yields the smallest sum SE for most values of $\taudmrs$. All curves rise with $\taudmrs$ until some value, and then decrease again. When $\taudmrs$ is too small, association and clustering are inefficient, as some UEs can possibly not connect to RUs with significant channel gain since all DMRS pilots at that RU are already assigned. When $\taudmrs$ is too large, pilot redundancy becomes a limiting factor of the system performance and some RUs may serve $| \Uc_\ell | < \taudmrs$ UEs. As $\tau_p$ is the maximum number of UEs that each RU can possibly serve, the system configuration with $L=10$ achieves the maximal sum SE for larger $\taudmrs$ compared to $L=20$ and $L=40$. With a more distributed antenna configuration, each RU covers a smaller region, and the expected number of UEs in its coverage area is lower. Thus, a smaller DMRS pilot dimension is required to serve all UEs with significant channel gain in the coverage area.

Note that the UL Tx power $P^{\rm ue}_{\rm tx}$ (and hence also the parameter $\SNR$) increases as $L$ decreases (see Table \ref{table_noise_snr}). This is due to 
our constraint that the Tx power should be sufficient to achieve a certain range $d_L$ and in order to effectively allow the cluster formation. 
The array beamforming gain (increasing with $M$) does not compensate for the larger distance between RUs (decreasing with $L$) for a constant product $LM$ of total number of 
system antennas. Note also that because of UL-DL duality, 
	the overall DL Tx power of all RUs is balanced with the UL Tx power.
	This confirms the fact that, as already noticed in several works (e.g., see \cite{9336188} and references therein) more distributed antenna configurations 
have the potential to yield both higher spectral and energy efficiency. Of course, the downside is that they require a larger number of RU sites and a corresponding denser
fronthaul network to connect them with the DUs. 


\section{Conclusions}  \label{sec:conclusions}

In this paper we considered a scalable cell-free user-centric wireless network architecture and introduced UL linear receive schemes based on 
cluster-level zero-forcing and local (per-RU) MMSE combining with cluster-level combining. A novel UL-DL duality based on the concept of ``nominal'' SINR was introduced, 
which is a useful proxy for the actual SINR experienced by the user receivers, but it can be computed at the cluster processors based solely on the 
CSI that can be effectively acquired via local UL DMRS pilots. This allows the reuse of the UL receive vectors as DL precoders, achieving virtually balanced UL-DL user rates. 
Of course, the imbalance between UL and DL traffic demands can be handled by allocating a different number of slots for the UL and DL in the TDD frame, however 
such scheduling aspects are out of the scope of this paper. 
We formulated the proposed receive/precoding schemes in the assumption of ideal partial CSI. Then, we used a plug-in approach and simply used the 
estimated channels via UL DMRS pilots for the actual performance. We proposed a channel subspace projection method that exploits the knowledge of 
the user channel's dominant subspace (which can be quite low-dimensional in semi-LOS propagation conditions typical of short ranges in dense networks) in order to 
partially but significantly eliminate DMRS pilot contamination. Then, we proposed  a novel UL SRS pilot scheme using orthogonal latin squares frequency hopping and 
a channel subspace estimation scheme based on R-PCA. A small number of strong co-pilot users is the result of the frequency hopping, and the R-PCA eliminates the remaining outliers containing heavily contaminated samples. The obtained subspace estimates are used for instantaneous channel estimation based on subspace projection, yielding an overall scheme for scalable operations in cell-free user-centric networks, including the estimation of long-term channel statistics (in this case, the channel subspaces), which has been 
often assumed to be known in previous literature, without providing a concrete and effective learning scheme. In this sense, our paper fills this gap. 
For the system simulations, we proposed to use the conditional distribution of the estimated subspace projected on DFT columns, where the conditioning variable is the channel gain, 
instead of running the R-PCA algorithm at each RU for all associated UEs. This method is validated through simulations and greatly reduces the computational complexity of simulating cell-free user-centric networks with possibly thousands of RU-UE pairs. 
The simulation results show that instantaneous channel estimation based on R-PCA subspace estimates can approach the system performance in the idealized cases of perfect subspace knowledge and ideal partial CSI, respectively. This essentially solves the problem of pilot contamination in cell-free TDD wireless networks. 


\appendices

	\section{On the accuracy of OER and UatF} \label{OER-UatF}
	
	The UatF lower bound to the achievable ergodic rate of the single-user channel
	``seen'' at the receiver of cluster $\Cc_k$, given in \eqref{received-UL}, 
	is given by \cite{Larsson-book,9336188} 
	\begin{equation} 
		R_k^{\rm UatF} = \log \left ( 1 + \frac{|\EE[ g_{k,k} ]|^2}{\SNR^{-1} + {\rm Var}(g_{k,k}) + \sum_{j \neq k} \EE[ |g_{k,j}|^2] } \right ),  \label{UatF}
	\end{equation}
	where we define the coefficients $g_{k,j} = \vvv_k^\herm \hh_j$ for the sake of notation simplicity. 
	In practical systems, it is usual to include {\em dedicated pilot symbols} in the payload (i.e., the precoded data packets)
	in order to enable coherent detection. In this Appendix, we develop a lower bound on the ergodic achievable rate 
	which can be regarded as a conditional version of the UatF bound, given the additional observation of {\em one} dedicated pilot per block. 
	Notice that a conditional UatF bound is already derived in \cite{Larsson-book,9336188} in full generality. Nevertheless, 
	it is useful to derive the bound directly and explicitly in this specific case, to compare the goodness of OER and UatF to predict the actual achievable system 
	performance for the system at hand. 
	
	Let the coherence block length be $T$ dimensions.  The channel seen at the receiver of cluster $\Cc_k$ is given by 
	$y_k[i] = g_{k,k} s_k[i] + e_k[i] + z_k[i]$, for $i = 1, \ldots, T-1$, where  $e_k[i]$ and $z_k[i]$ denote  interference and noise, respectively, plus an additional 
	channel observation of the form
	$$
	y_k^{\rm pilot} = g_{k,k} p_k  + e_k + z_k
	$$
	in correspondence of a single dedicated pilot symbol  $p_k$, included in the transmitted data slot.
	The relevant mutual information per block is given by 
	\begin{equation} 
		I(s_k[1:T-1] ; y_k[1:T-1], y_k^{\rm pilot})  =  I(s_k[1:T-1] ; y_k[1:T-1] | y_k^{\rm pilot}), \label{Info-block}
	\end{equation}
	since the sequence of symbols $s_k[1:T-1] = ( s_k[1], \dots,  s_k[T-1] )$ (same notation is used for $y_k[1:T-1]$) is independent 
	of $y_k^{\rm pilot}$. 
	
	
	Now, replicating the derivation of the UatF bound (see Lemma 2 in \cite{8304782}) and assuming that $s_k[i]$ are i.i.d. $\sim \Cc\Nc(0,\Ec_s)$, 
	the mutual information per block can be lower bounded as
	\begin{eqnarray}
		&& I(s_k[1:T-1] ; y_k[1:T-1] | y_k^{\rm pilot} ) \\
		& &= h(s_k[1:T-1] | y_k^{\rm pilot}) - h(s_k[1:T-1] | y_k[1:T-1] , y_k^{\rm pilot}) \\
		& & = h(s_k[1:T-1] ) - h(s_k[1:T-1] | y_k[1:T-1] , y_k^{\rm pilot}) \label{eq:mutual_I_s_p_independent} \\
		& &= h(s_k[1:T-1] ) - h(s_k[1:T-1] - \widehat{s}_k[1:T-1] | y_k[1:T-1] , y_k^{\rm pilot}) \\
		& &\geq (T-1) \log (\pi e \Ec_s) - (T - 1) \log ( \pi e \; \MMSE(S|Y, Y^{\rm pilot})) \label{eq:mutual_I_mmse},
	\end{eqnarray}
	where (\ref{eq:mutual_I_s_p_independent}) follows from the independence of $s_k([1:T-1])$ and $y_k^{\rm pilot}$, and where
	$\MMSE(S|Y, Y^{\rm pilot})$ is the MMSE resulting from the estimation of $S$ from $(Y, Y^{\rm pilot})$ in the observation model
	\begin{equation}
		\left \{ 
		\begin{array}{l}	Y = g_{k,k} S + W, \\
			Y^{\rm pilot} = g_{k,k} \sqrt{\Ec_p} + W'. 
		\end{array}
		\right .   \label{Y-obs}
	\end{equation}
	Here, $\Ec_p = |p_k|^2$ is the energy per symbol of the dedicated pilot, $S$ is distributed as the symbols $s_k[i]$, 
	$W$ and $W'$ are mutually independent Gaussian random variables with mean zero and same variance as the interference plus noise 
	in \eqref{received-UL}. 
	Imposing that the average energy per symbol in each slot is 1, we find that $\Ec_s$ and $\Ec_p$ are related by
	\begin{equation} 
		\Ec_s = (T - \Ec_p)/(T-1).  \label{energy-per-symbol}
	\end{equation}
The dedicated pilot symbol is placed in the slot according to some known 
pseudo-random assignment, proper of each user and given by some some protocol. Hence, 
by (\ref{energy-per-symbol}) the average interfering energy per symbol is equal to 1. This results in
	\begin{equation} 
		\EE[|W|^2] = \EE[|W'|^2] = \SNR^{-1} + \sum_{j \neq k} \EE[ |g_{k,j}|^2].  \label{varW}  
	\end{equation}
	Any suboptimal estimator yields a mean squared error (MSE) larger than $\MMSE(S|Y, Y^{\rm pilot})$ and therefore a lower bound for (\ref{eq:mutual_I_mmse}).
	Here, we consider a ``quasi-coherent'' scheme that estimates $g_{k,k}$ from $Y^{\rm pilot}$ and then uses it to estimate $S$ from $Y$. 
	Letting $\bar{g}_{k,k} = \EE[ g_{k,k}]$, we use the linear MMSE estimator 
	\begin{eqnarray}
		\hat{g}_{k,k}  =  \bar{g}_{k,k} + \frac{{\rm Var}(g_{k,k}) \sqrt{\Ec_p} }{{\rm Var}(Y^{\rm pilot}) } (Y^{\rm pilot} - \bar{g}_{k,k} \sqrt{\Ec_p}) \label{eq:g_hat}.
	\end{eqnarray} 
	The resulting MSE  is given by
	\begin{align}
		\MMSE (g_{k,k} | Y^{\rm pilot}) & = \EE[ | g_{k,k} - \hat{g}_{k,k} |^2 ]  = \frac{{\rm Var}(W') {\rm Var}(g_{k,k})}{{\rm Var}(g_{k,k}) \Ec_p + {\rm Var}(W')}. \label{MMSEgkk}
	\end{align}
	Now, we rewrite the first line of (\ref{Y-obs}) as
	\begin{eqnarray}
		Y =  \hat{g}_{k,k} S + \underbrace{(g_{k,k} - \hat{g}_{k,k} ) S + W}_{\widetilde{W}}.  \label{Y-obs1}
	\end{eqnarray}
	Notice that $\widetilde{W}$ and $\hat{g}_{k,k} S$ are uncorrelated since, by the property of the linear MMSE estimator, we have that 
	$\EE[ \hat{g}_{k,k} ( g_{k,k} - \hat{g}_{k,k})^* | Y^{\rm pilot}] = 0$. Hence, using the law of iterated expectation, we have
	\[ \EE[ \hat{g}_{k,k} S \widetilde{W}^*] =  \EE[ \EE[ \hat{g}_{k,k} S ((g_{k,k} - \hat{g}_{k,k})S + W)^* | Y^{\rm pilot}] ]   =  0. \]
An upper bound on $\MMSE(S | Y, Y^{\rm pilot})$ is obtained by the MMSE estimator of $S$ given $Y$ in (\ref{Y-obs1}), 
	by treating  $\widetilde{W} = (g_{k,k} - \hat{g}_{k,k} ) S + W$ as an {\em uncorrelated additive noise} with variance 
	\begin{equation} 
		{\rm Var}(\widetilde{W}) = \EE[ |\widetilde{W}|^2] = \Ec_s \MMSE(g_{k,k} |Y^{\rm pilot}) + {\rm Var}(W). \label{VarWtilde}
	\end{equation}
	This yields immediately the MMSE upper bound
	\begin{equation}
		\MMSE(S|Y, Y^{\rm pilot}) \leq \EE \left [ \frac{\Ec_s {\rm Var}(\widetilde{W})}{|\hat{g}_{k,k}|^2 \Ec_s + {\rm Var}(\widetilde{W})} \right ] \label{MMSE-upperbound}
	\end{equation}
	Using \eqref{MMSE-upperbound} in \eqref{eq:mutual_I_mmse} and using the monotonicity of the logarithm, we obtain
	\begin{equation}
		I(s_k[1:T-1] ; y_k[1:T-1] | y_k^{\rm pilot} ) \geq (T-1) \left ( \log \Ec_s  - \log \left ( \EE \left [ \frac{\Ec_s {\rm Var}(\widetilde{W})}{|\hat{g}_{k,k}|^2 \Ec_s + {\rm Var}(\widetilde{W})} \right ] 
		\right ) \right ). 
	\end{equation}
	Then, since $- \log$ is a convex function, applying Jensen's inequality and using (\ref{VarWtilde}), (\ref{MMSEgkk}), and (\ref{varW}),  we obtain
	\begin{align}
		& I(s_k[1:T-1] ; y_k[1:T-1] | y_k^{\rm pilot} ) \nonumber \\
		& \geq (T-1)  \EE \left [ \log \left ( 1 + \frac{|\hat{g}_{k,k}|^2 \Ec_s}{{\rm Var}(\widetilde{W})} \right ) \right ] \nonumber \\
		& = (T-1) \EE \left [ \log \left ( 1 + \frac{|\hat{g}_{k,k}|^2 \Ec_s}{\SNR^{-1} + \EE[ | g_{k,k} - \hat{g}_{k,k} |^2 ] \Ec_s  + \sum_{j \neq k} \EE[ |g_{k,j}|^2]} \right ) \right ]. 
		\label{final-bound}
	\end{align}
	The desired achievable rate lower bound expressed in bits per channel use is eventually obtained by dividing the expression in the RHS of (\ref{final-bound}) by $T$. 
	By comparing this with the OER in (\ref{ergodic_rate_ul})  we notice the following differences: i) a factor $(1-1/T)$ due to the one dedicated pilot per block redundancy; ii) 
	the useful signal term $g_{k,k}$ in the SINR numerator in (\ref{ergodic_rate_ul}) is replaced by the estimated useful signal term $\hat{g}_{k,k}$; iii) 
	a ``self-interference'' term $\EE[ | g_{k,k} - \hat{g}_{k,k} |^2 ] \Ec_s$ due to the non-perfect knowledge of the useful signal term appears in the SINR denominator of \eqref{final-bound}; iv) the energy per symbol (which is equal to 1 in (\ref{ergodic_rate_ul})) is replaced by $\Ec_s \leq 1$, assuming that energy $\Ec_p \geq 1$ is allocated to the dedicated pilot.


	\begin{figure}[t]
		\centerline{
			\includegraphics[width=.5\linewidth]{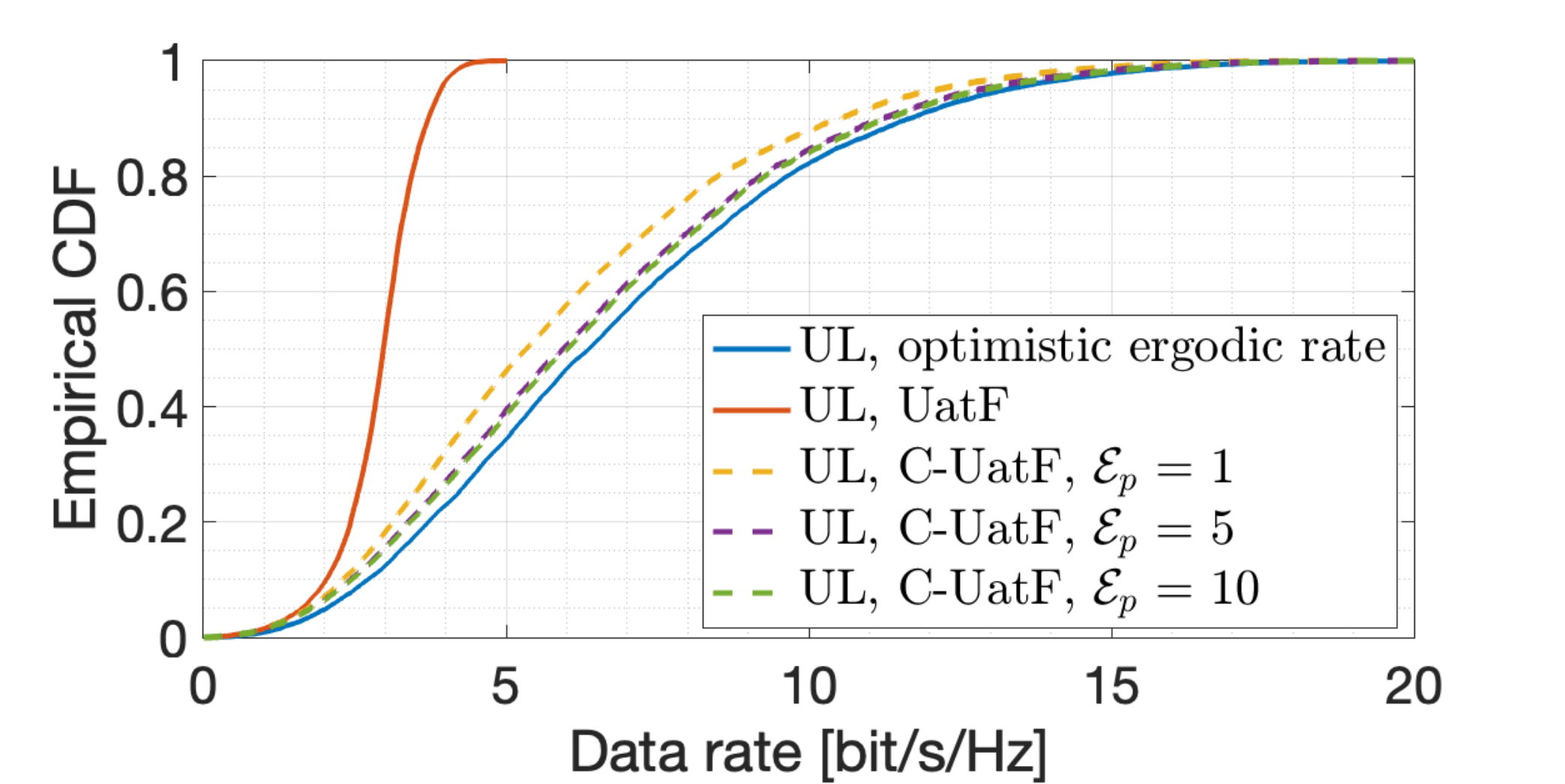} }
		\vspace{-.6cm}
		\caption{The ergodic rate bounds using LMMSE receivers with cluster-level combining and subspace projection channel estimation.}
		\vspace{-.8cm}
		\label{fig:uatf_optimistic_erg_rate}
	\end{figure}
	
	We refer to the achievable ergodic rate lower bound obtained in this section as the ``Conditional UatF'' (C-UatF) bound, since it can be seen as a conditional version of the 
	UatF bound given the dedicated pilot observation per block.  Fig. \ref{fig:uatf_optimistic_erg_rate} shows the ergodic rate bounds for a system with the same parameters as the one from Fig. \ref{cdfplots_UL_DL_L40_pca_vs_sp_ideal}. We evaluate different values of $\Ec_p$ and observe that the gap between the OER 
	and the C-UatF bounds can be made very small by an appropriate choice of $\Ec_p$. In practice, pilot/data power imbalance is possible and does not entail a large 
	peak-to-average ratio problem since a real system uses OFDM, and the pilot per RB is inserted in the frequency domain, and therefore it is smeared out in the time domain 
	after the OFDM modulator. This shows that for systems as those studied in this paper, with a relatively small number of antennas per RU and significant antenna correlation due to the restricted scattering angular spread, OER is a much better predictor of the rates effectively achievable with a reasonably designed quasi-coherent scheme than the ubiquitous UatF bound, which may be over-conservative.

\fontsize{10}{12}\selectfont
\bibliographystyle{IEEEtran}
\bibliography{massive-MIMO-references}

\end{document}